\documentclass[aps,prl,a4paper,twocolumn,superscriptaddress,nobibnotes]{revtex4-1}
\usepackage{helvet}
\usepackage{graphicx}
\usepackage{xspace}
\usepackage{amsmath,amssymb,bm}
\usepackage[usenames,dvipsnames]{xcolor}

\usepackage{mathrsfs}

\usepackage{siunitx}

\newif\ifdrafttext
\drafttextfalse 
\ifdrafttext \usepackage[colorlinks,urlcolor=black,citecolor=black,linkcolor=black]{hyperref} \else   \usepackage[hidelinks]{hyperref} \fi

\usepackage[all]{hypcap}

\newcommand{\QCDaffiliation}{QCD Labs, QTF Centre of Excellence, Department of Applied Physics, Aalto University, P.O. Box 13500, FI-00076 Aalto, Finland}
\newcommand{\NISTaffiliation}{National Institute of Standards and Technology, Boulder, Colorado, 80305, USA}
\newcommand{\VTTaffiliation}{VTT Technical Research Centre of Finland Ltd, P.O. Box 1000, FI-02044 VTT, Finland}

\newcommand{\OULUaffiliation}{Research Unit of Nano and Molecular Systems, University of Oulu, P.O. Box 3000, FI-90014 Oulu, Finland}

\begin{document}

\title{Optimized heat transfer at exceptional points in quantum circuits}

\author{M. Partanen}
\affiliation{\QCDaffiliation}
\author{ J. Goetz}
\affiliation{\QCDaffiliation}
\author{ K. Y. Tan}
\affiliation{\QCDaffiliation}
\author{ K. Kohvakka}
\affiliation{\QCDaffiliation}
\author{ V. Sevriuk}
\affiliation{\QCDaffiliation}
\author{ R. E. Lake}
\affiliation{\QCDaffiliation}
\affiliation{\NISTaffiliation}
\author{ R. Kokkoniemi}
\affiliation{\QCDaffiliation}
\author{ J. Ikonen}
\affiliation{\QCDaffiliation}
\author{ D. Hazra}
\affiliation{\QCDaffiliation}
\author{ A. M\"akinen}
\affiliation{\QCDaffiliation}
\author{ E. Hyypp\"a}
\affiliation{\QCDaffiliation}
\author{ L. Gr\"onberg}
\affiliation{\VTTaffiliation}
\author{ V. Vesterinen}
\affiliation{\QCDaffiliation}
\affiliation{\VTTaffiliation}
\author{ M. P. Silveri}
\affiliation{\QCDaffiliation}
\affiliation{\OULUaffiliation}
\author{ M. M\"ott\"onen}
\affiliation{\QCDaffiliation}

\date{December 6, 2018}

\begin{abstract}

Superconducting quantum circuits are potential candidates to realize a large-scale quantum computer. 
The envisioned large density of integrated components, however, requires a proper thermal management and control of dissipation. 
To this end, it is advantageous to utilize tunable dissipation channels and to exploit the optimized heat flow at exceptional points (EPs).
Here, we experimentally realize an EP in a superconducting microwave circuit consisting of two resonators. 
The EP is a singularity point of the Hamiltonian, and corresponds to the most efficient heat transfer between the resonators without oscillation of energy.
We observe a crossover from underdamped to overdamped coupling via the EP by utilizing photon-assisted tunneling as an \emph{in situ} tunable dissipative element in one of the resonators.
The methods studied here can  be applied to different circuits to obtain fast dissipation, for example, for initializing qubits to their ground states.
In addition, these results pave the way towards thorough investigation of parity--time ($\mathcal{PT}$) symmetric systems and the spontaneous symmetry breaking in superconducting microwave circuits operating at the level of  single energy quanta.

\end{abstract}

\maketitle

\renewcommand{\figurename}{Figure}
\renewcommand{\tablename}{Table}

Systems with effective non-Hermitian Hamiltonians have been actively studied in various setups in recent years~\cite{Milburn_2015,Xu_2016,Doppler_short_2016,Mandal_2016,Ding_2016,Ding_2018,SanJose_2016}.
They show  many intriguing properties  such as singularities in their energy spectra~\cite{Kato_1966,Dembowski_2001,Heiss_2004, Berry_2004, Heiss_2012,Cao_2015}.
A square-root singularity point in the parameter space of a non-Hermitian matrix is called an exceptional point, EP, if the eigenvalues coalesce~\cite{Heiss_2012,Cao_2015}.
Previously, EPs have been shown to emerge, for example, in non-superconducting microwave circuits, laser physics, quantum phase transitions, and atomic and molecular physics~\cite{Heiss_2012,Cao_2015}.
The fascinating effects of EPs include the disappearance of the beating Rabi oscillations~\cite{Dietz_2007}, chiral states in microwave systems~\cite{Dembowski_2003},  and spontaneous symmetry breaking in systems with parity- and time-reversal  ($\mathcal{PT}$)  symmetry~\cite{Bender_1998,Guo_short_2009,Chtchelkatchev_2012,ElGanainy_2018}.
In the quantum regime, $\mathcal{PT}$-symmetric systems may show features that are different from the semiclassical predictions, such as new phases owing to quantum fluctuations~\cite{Kepesidis_2016,ElGanainy_2018}.
Despite the active research on EPs, they have not been thoroughly investigated in superconducting microwave circuits to date~\cite{Quijandria_2018}.

Superconducting microwave circuits provide an ideal platform to realize  various quantum devices, such as ultra\-sensitive photon detectors and counters~\cite{Inomata_2016,Govenius_2016,Ding_short_2017,Opremcak_short_2018}, and potentially even a large-scale quantum computer~\cite{Ladd,Clarke}, or a quantum simulator~\cite{Georgescu_2014} in the framework of circuit quantum electrodynamics~\cite{Blais_2004,Wallraff_short_2004}.
Notably, superconducting qubits have been shown to approach the required coherence times~\cite{Barends_short_2014,Kelly_short_2015} for quantum error correction~\cite{Lidar_book_2013,Terhal_2015}.
However, despite the tremendous interest in superconducting microwave circuits in  recent years~\cite{You_2011,Devoret_2013,Wendin_2017},  there are still many issues to be solved before a fully functional quantum computer is possible.
For example, the precise engineering of energy flows between different parts of the circuit in scalable architectures is of utmost importance since unwanted heat is a typical source of decoherence in qubits~\cite{Clerk_2007,Goetz_short_2017}.
In many error correction codes, qubits are repeatedly initialized, which requires fast and efficient cooling schemes~\cite{Geerlings_short_2013,Bultink_short_2016,Wong_2018}.
One promising method for absorbing energy and initializing qubits to their ground state is based on resonators with tunable dissipation~\cite{Tuorila_2017}.

The recently developed quantum-circuit refrigerator (QCR)~\cite{Tan_2017, Masuda_short_2018} provides great potential for both qubit initialization and thermal management since it enables tunability of energy dissipation rates over several orders of magnitude in a superconducting microwave resonator~\cite{Silveri_short_2018}. 
Operation of the QCR relies on inelastic tunneling of electrons through a normal-metal--insulator--superconductor (NIS) junction~\cite{Silveri_2017}.
The tunneling electrons can absorb or emit photons to a resonator which allows to control the coupling strength to a low-temperature bath \emph{in situ}.
This tunable coupling strength has also been shown to induce a broadband Lamb shift~\cite{Silveri_short_2018}.
Furthermore, elastic tunneling can be utilized for temperature control of the normal-metal electrons~\cite{Nahum_1994,Leivo_1996, Giazotto}, and for precise thermometry down to millikelvin temperatures~\cite{Feshchenko_short_2015,Giazotto}.
Recently, NIS junctions have also been utilized in a realization of  a quantum heat valve~\cite{Ronzani_2018}, and phase-coherent caloritronics~\cite{Fornieri_2017}.

In this work, we combine the advantages of tunable dissipation and EPs to optimize the heat flow in a superconducting microwave circuit. 
To this end, we investigate a circuit consisting of two coupled resonators, one of which is equipped with NIS junctions and a flux-tunable resonance frequency (Fig. 1). 
We denote the NIS junctions and the normal-metal island that is capacitively coupled to the resonator as a QCR. 
Thanks to the voltage-tunable dissipation within the QCR and the flux-tunable resonance frequency, an EP arises in the Hamiltonian that describes the modes of the coupled resonator system. 
We investigate the emergence of the EP using frequency and dissipation as control parameters (Fig. 2) and verify its properties experimentally by measuring the microwave transmission coefficient (Fig. 3).
The optimal heat flow given by the coupling strength can be reached at the EP (Fig.~\ref{fig:transition_rates}).
Different types of tunable resonators have been studied in recent years~\cite{Wang_2013,Pierre_2014,Baust_short_2015, Vissers_2015, Wulschner_short_2016, Adamyan_2016,Pierre_2018, Wong_2018, Partanen_short_2018} but not with voltage-tunable dissipation.
Our work demonstrates a platform to control the local heat transport between neighboring nodes in a quantum electrical circuit.  
In addition to thermal management within superconducting multi-qubit systems, these methods may be applicable to thermally assisted quantum annealing~\cite{Dickson_short_2013} and to studies of the eigenstate thermalization hypothesis in many-body quantum problems~\cite{Nandkishore_2015}. 
Furthermore, our work is an important step towards the investigation of $\mathcal{PT}$-symmetric systems at the quantum level that can be realized with circuit quantum electrodynamics architectures~\cite{Metelmann_2018,
Quijandria_2018}.

\section*{Results}

\subsection*{Experimental samples}

\begin{figure}[t]
\centering
\includegraphics[width=85mm]{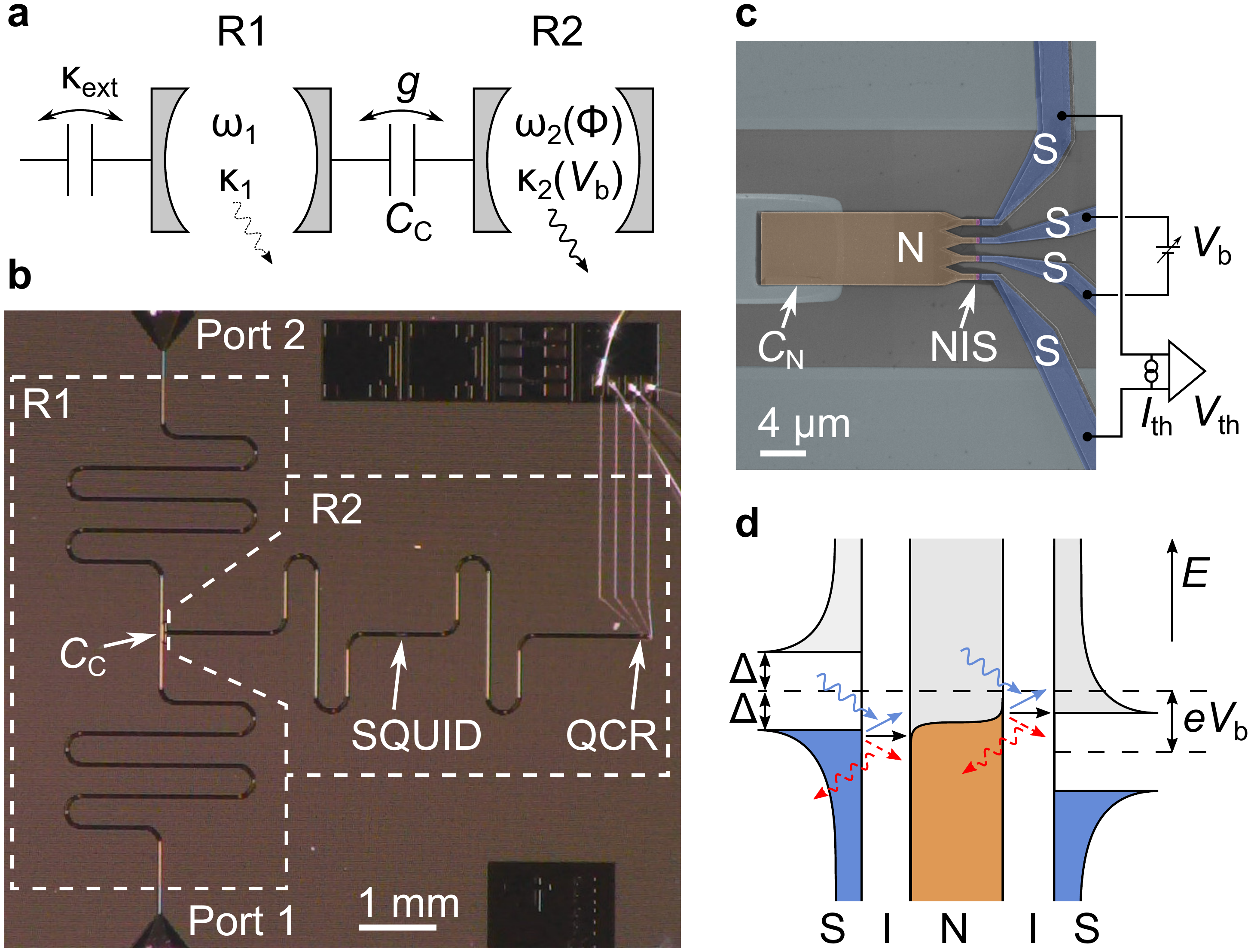}
\caption{ Sample structure.
(a)~The sample consists of two capacitively coupled resonators, R1 and R2, which are presented as analogous cavities with coupling strength $g$. 
The  primary resonator R1 has a fixed dissipation rate $\kappa_1$ and angular frequency $\omega_1$  whereas the dissipative resonator R2 has a tunable  dissipation $\kappa_2(V_\textrm{b})$ controlled by a QCR, and angular frequency $\omega_2(\Phi)$ tuned with a SQUID. 
The coupling strength to external ports is denoted by $\kappa_\textrm{ext}$.
(b)~Optical micrograph of the sample. The transmission coefficient $S_{21}$ is measured from Port~1 to Port~2.
(c)~False-colour scanning electron micrograph of the QCR together with a schematic control circuit.
The QCR consists of normal-metal (N) and superconducting (S) electrodes separated by an insulator (I).
The QCR is operated with bias voltage $V_\textrm{b}$, and the electron temperature of the normal metal is obtained from voltage $V_\textrm{th}$ and current $I_\textrm{th}$.
The micrographs in (b) and (c) are from Sample~B.
(d)~The operation principle of a SINIS junction.  
The occupied states in the  superconductor density of states are shown in blue, the occupation of the normal metal is given by the Fermi distribution shown in orange, and the empty states are shown in gray with energy $E$ on the vertical axis.
The Fermi levels of the superconducting electrodes (dashed lines) are shifted by applying a voltage $V_\textrm{b}$.
The black arrows indicate elastic tunneling, and blue arrows inelastic tunneling with photon absorption. 
The red dashed arrows show photon emission that is suppressed due to lack of available unoccupied states on the other side of the tunneling barrier.
}
\label{fig:sample_structure}
\end{figure}

Our samples consist of two coplanar waveguide resonators, R1 and R2, which are capacitively coupled to each other, as depicted in Fig.~\ref{fig:sample_structure}(a), (b) (see also Supplementary Fig.~\ref{fig:sample_details}). 
The resonator R1 has a fixed fundamental   frequency at $\SI{2.6}{\giga\hertz}$. 
This mode does not couple strongly to the resonator R2 owing to the voltage node in the middle of the resonator R1 where the coupling capacitor $C_\textrm{C}$  is located. 
Therefore, we focus on the first excited mode of R1, with frequency $f_1 = \omega_{1}/(2\pi) = \SI{5.2}{\giga\hertz}$, which has a voltage antinode at the coupling point of the resonators. 
The resulting capacitive coupling between the resonators has a strength $g/(2\pi) = \SI{7.2}{\mega\hertz}$.
In contrast to R1, the resonator R2 has a flux-tunable resonance frequency $\omega_{2}(\Phi)$ owing to a superconducting quantum interference device (SQUID), and a voltage tunable loss rate $\kappa_{2}(V_\textrm{b})$ owing to the QCR.
Here, $\Phi$ and $V_\textrm{b}$ are the magnetic flux applied to the SQUID loop and the voltage bias of the QCR, respectively.
The inductance of the SQUID and, hence, also the resonance frequency of R2 are periodic in flux with a period of the flux quantum $\Phi_0 = e/(2h)$.
Consequently, due to the coupling of the resonators,  R1 also shows flux-dependent features.
We show the QCR in Fig.~\ref{fig:sample_structure}(c) and schematically present its operation principle    in Fig.~\ref{fig:sample_structure}(d).
The difference in photon absorption and emission rates originates from the gap of $2\Delta$ in the density of states of the superconductor, and the difference can be utilized to cool down quantum circuits~\cite{Tan_2017,Silveri_2017}.
The sample fabrication is described in Methods.
We study two samples with different R2 resonator lengths, Sample~A ($\SI{12}{mm}$) and Sample B ($\SI{13}{mm}$).
The R1 resonator has a length of $\SI{24}{\milli\meter}$ in both samples.
Sample parameters are summarized in Supplementary Table~\ref{tab:simulation_parameters}.

\subsection*{Exceptional points}

\begin{figure}[t]
\centering
\includegraphics[width=75mm]{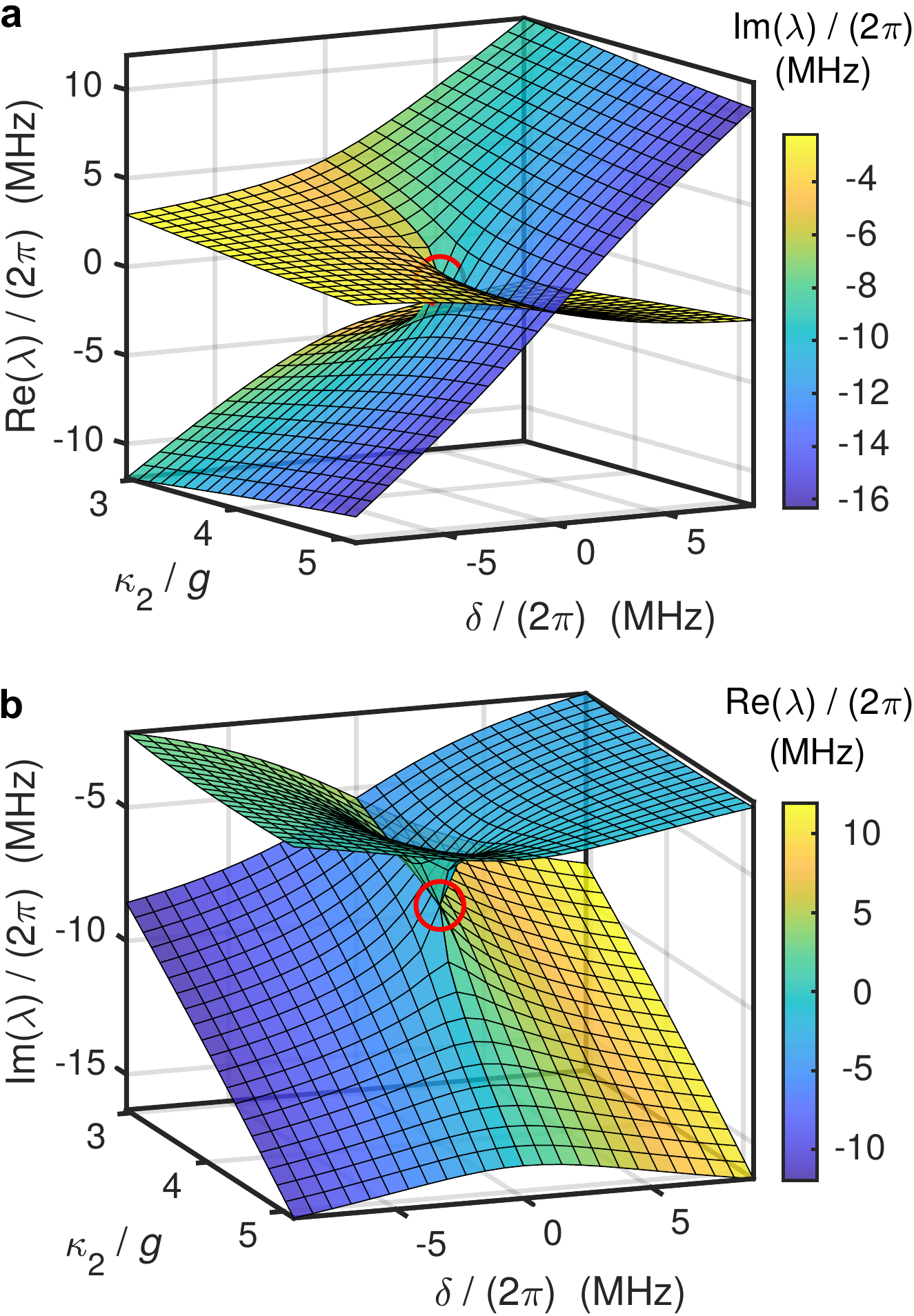}
\caption{ Eigenvalues of the effective Hamiltonian and the exceptional point.
Calculated (a)~mode frequency shifts $\textrm{Re}(\lambda)$ with respect to the uncoupled mode frequency of R1, and (b)~negative mode decay rates $\textrm{Im}(\lambda)$ as functions of the decay rate $\kappa_2$  and frequency detuning $\delta$.
The figure shows both $\lambda_+$ and $\lambda_-$ calculated according to Eq.~\eqref{eq:lambda} with the experimental coupling strength $g/(2\pi) = \SI{7.2}{\mega\hertz}$ and decay rate $\kappa_1/(2\pi) = \SI{260}{\kilo\hertz}$.
The EP is located approximately at $\kappa_2 = 4g$, and $\delta =0$ as indicated with the red circle.
The matching between the panels (a) and (b) can be seen from the colors which denote $\textrm{Im}(\lambda)$ in (a) and $\textrm{Re}(\lambda)$ in (b).
}
\label{fig:EP_simulation}
\end{figure}

To study exceptional points, we utilize two control parameters in the effective Hamiltonian of the system: we use the voltage-tunable dissipation rate $\kappa_{2}(V_\textrm{b})$ and the flux-tunable detuning between the resonators $\delta =  \omega_{2}(\Phi) - \omega_{1} $. 
We study the system consisting of the two resonators in a rotating frame with a frequency corresponding to the uncoupled mode frequency of R1, similarly as in Ref.~\cite{Doppler_short_2016}.
Thus, the excitations of the system can be described with the effective non-Hermitian Hamiltonian 
 in matrix form in the basis $\psi = (A,B)^\textrm{T}$ where $A$ and $B$ are field amplitudes in R1 and R2, respectively,
 (see Methods and Ref.~\cite{Pierre_2018})
\begin{equation}
 H =  
 \begin{pmatrix}
  -i\frac{\kappa_1}{2}   &  g  \\
  g   & \delta -i\frac{\kappa_2}{2} 
 \end{pmatrix} ,
\label{eq:H}
\end{equation}
where $\kappa_1$ is the decay rate of the resonator R1. 
The eigenvalues of $H$ can be written as 
\begin{equation}
 \lambda_\pm = \frac{1}{4} \left(2 \delta-i \kappa_1-i \kappa_2 \pm s \right),
\label{eq:lambda}
\end{equation}
and the corresponding eigenvectors are
\begin{equation}
\psi_\pm = \left(\frac{ -2 \delta-i \kappa_1+i \kappa_2 \pm s}{4 g},1 \right)^\textrm{T},
\end{equation}
where
\begin{equation}
 s = \sqrt{4 \delta^2+16 g^2 -  ( \kappa_2-\kappa_1)^2  - i 4  \delta (\kappa_2- \kappa_1)}.
\end{equation}
Thus, the eigenvalues and eigenvectors coalesce when the  square-root term $s$ vanishes resulting in an EP.
Consequently, there is only a single eigenvalue and, importantly, there is also only a single eigenvector.
The EP occurs at $|\kappa_2-\kappa_1|=4g$, and $\delta = 0$.
In the following, we assume that $\kappa_1 \ll \kappa_2$, which is valid for our samples, as discussed below.  
Consequently, the condition for the EP simplifies to $\kappa_2=4g$.

To visualize the system singularity, i.e., the EP, we show the real and imaginary parts of $\lambda_{\pm}$ in Fig.~\ref{fig:EP_simulation}. 
The eigenvalues form a self-intersecting Riemann surface in the parameter space of $\kappa_2$ and $\delta$.
The imaginary part corresponds to mode decay, and real part to mode frequency deviation from the uncoupled mode frequency of R1.
Our system consisting of the resonators R1 and R2 can be considered as a single damped harmonic oscillator, where the energy oscillates between the two resonators.
In the underdamped case, $\kappa_2 < 4g$, the modes have an equal decay rate at zero detuning, and there is an anti-crossing of the mode frequencies.
In contrast, in the overdamped case, $\kappa_2 > 4g$, there is an anticrossing in the mode decay rates as a function of the detuning, and the mode frequencies are equal at zero detuning.
Consequently, one of the modes remains lossy whereas the other one has a low decay rate at different detunings.

Let us  connect the meaning of this critical point to the efficiency of energy transfer between the two resonators. 
In terms of coupled dissipative systems, the EP separates the system between the overdamped and underdamped regime being the point of critical coupling. 
It follows from the dynamics of the coupled system that  at this point the energy is transfered between the two resonators optimally fast without back and forth oscillation~\cite{Pierre_2018,Wong_2018}. 
In particular, the rate of heat transfer at zero detuning is given by $\kappa_\textrm{eff} = -2 \textrm{Im}(\lambda_\pm) \approx \kappa_{2} [1 \mp  \textrm{Re}( \sqrt{1- (4g / \kappa_{2})^{2}} )]/2$. 
Here, the  branch with the upper signs corresponds to a mode located predominantly in the primary resonator R1, and the branch with the lower ones in the dissipative resonator R2 (Supplementary Fig.~\ref{fig:eigenvalues_eigenvectors}).
Consequently, by reaching the EP at $\kappa_2=4g$, we operate our sample at a point of optimally efficient and nonreciprocal heat transfer out of R1.

\subsection*{Experimental observations}

\begin{figure}[t]
\centering
\includegraphics[width=85mm]{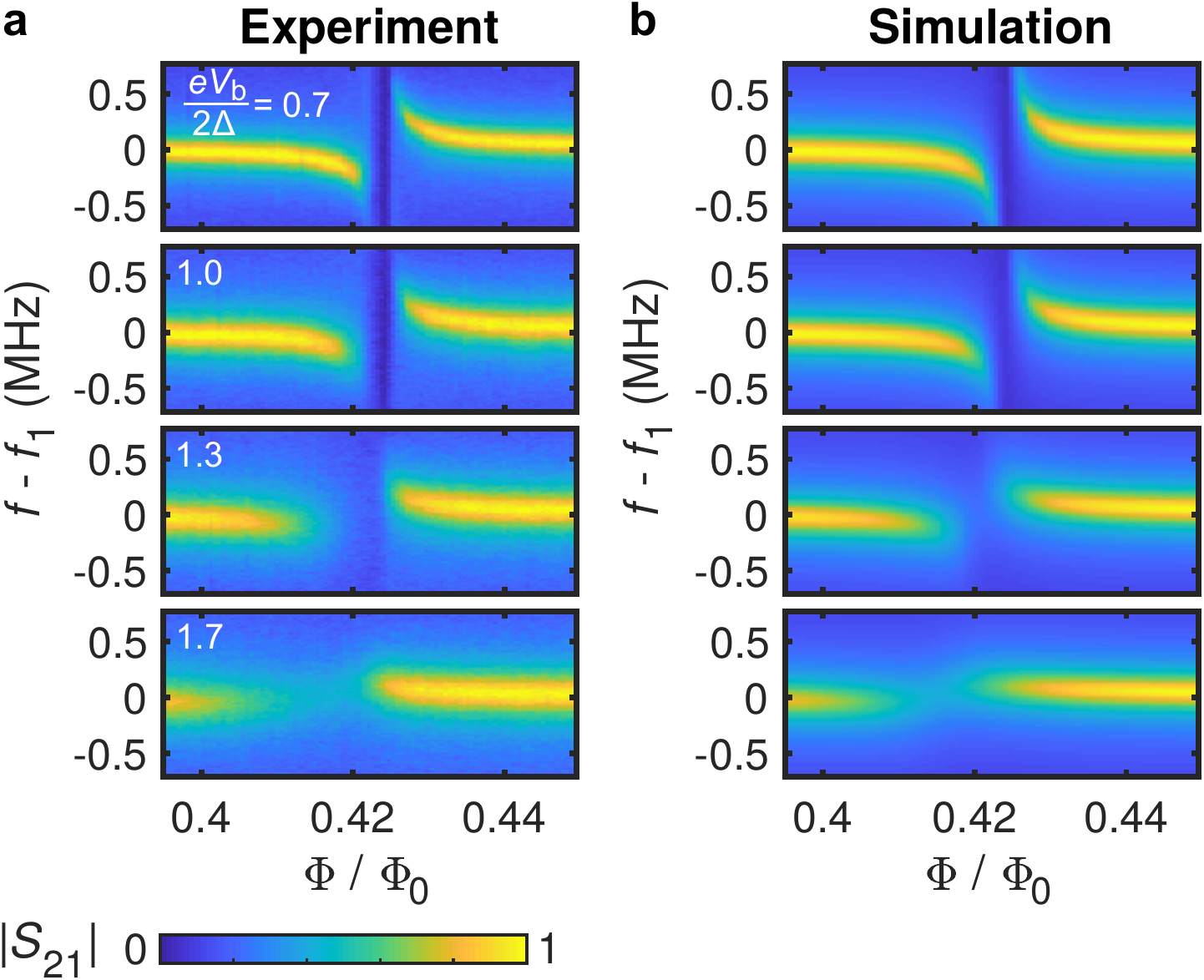}
\caption{ Scattering parameter of Sample~A.
(a)~Experimental and (b)~simulated transmission amplitudes as functions of frequency and magnetic flux through the SQUID.
The panels show the crossing of the second mode of R1 and the first mode of R2 at different bias voltages as indicated in the figure.
The EP is obtained at $eV_\textrm{b}/(2\Delta) \approx 1$, which approximately corresponds to the second panel from the top.
The maximum in each panel is normalized to unity, and the measured frequencies of R1 are shifted by the resonance frequency $f_1=5.223$~GHz, which is in  good agreement with the simulated frequency  $f_1 = 5.2$~GHz.
The input power at Port~1 is approximately $-100$~dBm.
}
\label{fig:meas_sim_crossing_selected}
\end{figure}

To explore the dissipative dynamics of the two coupled resonators, we measure the  flux- and frequency-dependent scattering parameter $S_{21}$ describing the transmission from Port~1 to Port~2 for different bias voltages using a vector network analyzer.
We tune the magnetic flux in a range where the frequency of R2 crosses that of R1.  
As  shown in Fig.~\ref{fig:meas_sim_crossing_selected}(a), we observe a transition from anticrossing to a single mode already indicating the presence of an EP in-between.
A broader range of bias voltages is shown in Supplementary Figs.~\ref{fig:meas_sim_crossing_sampleA} and \ref{fig:meas_sim_crossing_sampleB}.
To generate a quantitative description of our system, which is required for the investigation of EPs, we numerically simulate the scattering coefficient as shown in Fig.~\ref{fig:meas_sim_crossing_selected}(b). 
Here, we model the SQUID  as a flux-tunable inductor, and the QCR as an effective resistance $R_\textrm{eff}$ (see Methods and Supplementary Fig.~\ref{fig:measurement_setup}).
We extract $R_\textrm{eff}$ by fitting the circuit model to the experimental results, and use  $R_\textrm{eff}$ to obtain the damping rates of the dissipative resonator R2 as a function of the bias voltage (Methods).
The extracted resistances are given in Supplementary Fig.~\ref{fig:transition_rate_parameters}. 
The model  in Fig.~\ref{fig:meas_sim_crossing_selected} is in very good agreement with the experimental results.
In addition to $R_\textrm{eff}$, we also extract the coupling capacitance and the critical current of the SQUID from the simulation.
The coupling capacitance is found to be $\SI{3.8}{\femto\farad}$ which agrees  well with the finite-element-method simulation that yields approximately $\SI{5}{\femto\farad}$.

In Fig.~\ref{fig:meas_sim_crossing_selected}(a) the crossing of the modes as a function of the flux shifts slightly towards lower flux values.
We attribute this shift to heating of the SQUID, which results in a reduced critical current of the SQUID, and hence a larger inductance and lower resonance frequency of R2.
In principle, we vary also the Lamb shift~\cite{Silveri_short_2018}, which, however, causes only a minor effect since the resonance of R2 is very broad, and therefore, we neglect it in our model.
To verify our above assumption $\kappa_{1} \ll \kappa_{2}$, we measure the internal quality factor of the primary resonator R1  with R2  far detuned  at $\Phi =  \Phi_{0}/2$.
From the quality factor, we extract a loss rate $\kappa_{1}/(2\pi) \lesssim \SI{260}{\kilo\hertz}$ for both samples which is substantially lower than the extracted value of $\kappa_{2}$.  
We measure a slight temperature and power dependence of the quality factor of R1 as expected in the case of two-level fluctuators dominating the losses~\cite{Zmuidzinas_2012} (Supplementary Fig.~\ref{fig:meas_Qfactor}).

\begin{figure}[t]
\centering
\includegraphics[width=80mm]{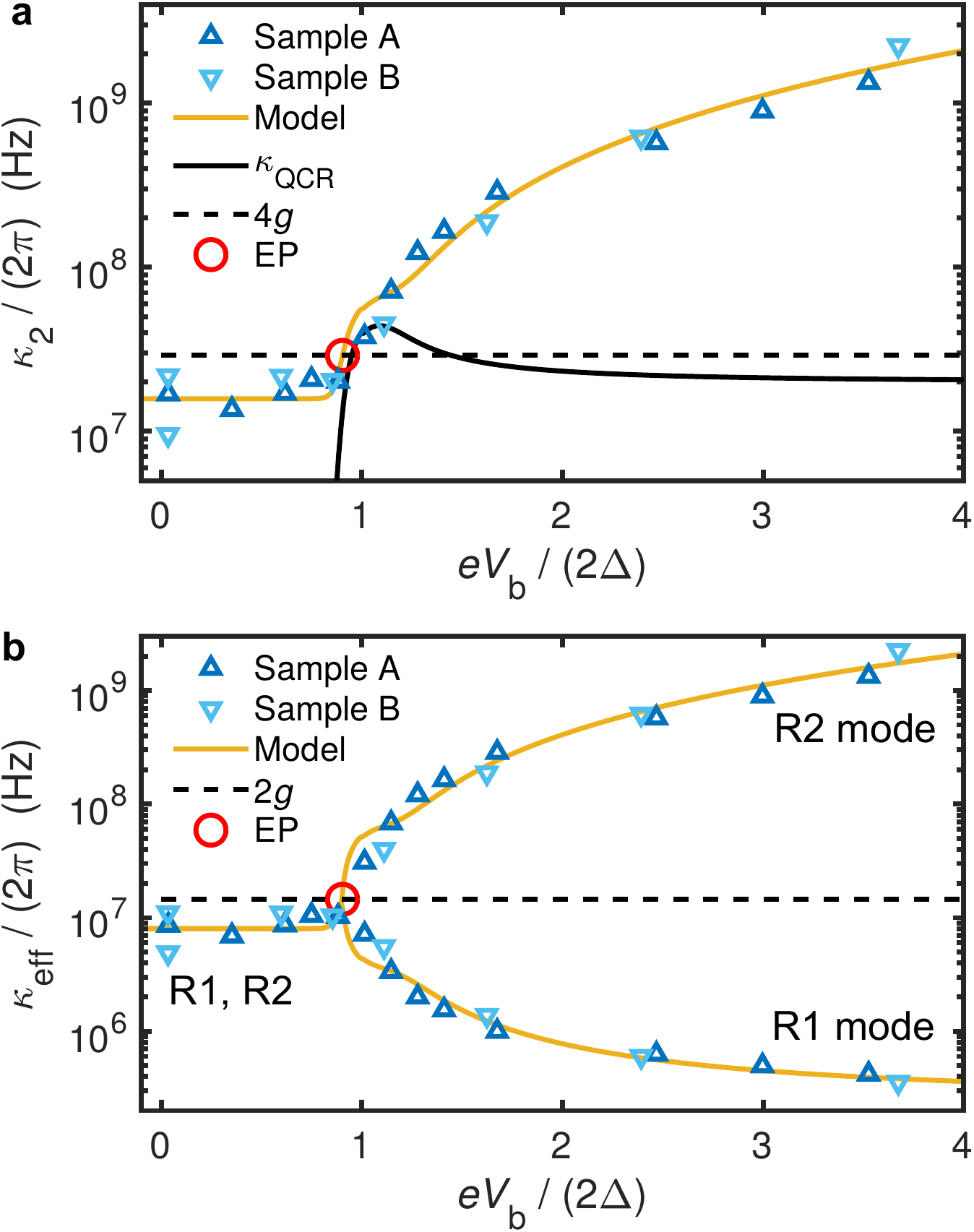}
\caption{ Transition rates.
(a)~Extracted decay rates of the bare resonator R2, $\kappa_\textrm{2}$,  for Samples A and B as  functions of the bias voltage  together with the full model (see Methods).
The EP is obtained at the intersection of $\kappa_2$ and the critical coupling  $4g$. 
Furthermore, the figure shows the theoretical coupling strength of the QCR,  $\kappa_\textrm{QCR}$, without taking dephasing and other voltage-dependent losses into account. 
The uncertainty of the data points is of the same order as the marker size.
(b)~Effective  decay rates of the coupled system  $\kappa_\textrm{eff} = -2 \textrm{Im}(\lambda_\pm)$ calculated from $\kappa_2$ at zero detuning using Eq.~\eqref{eq:lambda}.
The two branches at high voltages correspond to $\lambda_+$ and  $\lambda_-$ with the modes located predominantly in one of the resonators as indicated.
The maximum decay rate for R1 is obtained at the EP.
The damping rates of the modes are equal at $eV_\textrm{b}/(2\Delta)<1$ due to hybridization.
}
\label{fig:transition_rates}
\end{figure}

To demonstrate the presence of an EP, we show the extracted damping rates (Methods) of the bare resonator R2 in the absence of R1 in Fig.~\ref{fig:transition_rates}(a) as  functions of the bias voltage for Samples~A and B.
Both samples have similar damping rates  $\kappa_{1}$ and $\kappa_{2}$. 
The value of $\kappa_{2}$  can be tuned by approximately two orders of magnitude.
At low bias voltages $eV_\textrm{b}/(2\Delta)<1$, the rate $\kappa_2$ is below $4g$, and at  $eV_\textrm{b}/(2\Delta)> 1$ the damping rate exceeds $4g$.
We describe the origin of the tunable damping rates using a model that contains the photon absorption and emission at the QCR given by the rate $\kappa_\textrm{QCR}$, as well as constant internal losses $\kappa_\textrm{int,2}$, and voltage-dependent residual losses $\kappa_\textrm{r,2}$. 
We have designed our sample in such a way that $\kappa_\textrm{QCR}$ covers the critical damping rate, and thus the losses originating from the QCR are sufficient to realize the EP.
The damping rate $\kappa_\textrm{QCR}$ shown in Fig.~\ref{fig:transition_rates} is calculated using the measured electron temperatures of the normal-metal island (see Methods and Supplementary Fig.~\ref{fig:IV_temperature_curves}).
At higher voltages, the major contribution in $\kappa_2$ is given by the residual damping coefficient, as discussed in Methods.
This damping coefficient includes dephasing, quasiparticle losses, and resistive losses, and we extract its value based on the experimental rates.
The dephasing has a similar effect on the measured coherent photon population as photon absorption due to other loss mechanisms although pure dephasing does not reduce the total photon number in the resonators.

In Fig.~\ref{fig:transition_rates}(b) we show the damping rates of the coupled circuit calculated from $\kappa_2$ using Eq.~\eqref{eq:lambda} as $\kappa_\textrm{eff} = -2 \textrm{Im}(\lambda_\pm)$.
The maximum energy decay rate for the mode in R1 is obtained at the EP as discussed above, and it is given by  $\kappa_\textrm{max,1} =\max[ -2\textrm{Im}(\lambda_+)] \approx \kappa_2/2 = 2 g$. 
Thus, in the optimal case, the decay rate is limited by  the coupling strength between the resonators.

\section*{Discussion}

We have experimentally realized an exceptional point,  EP, in a superconducting microwave circuit.
We study the presence of the EP by observing a transition from an avoided crossing to single modulating resonance frequency.
At this point, we achieve a maximum heat transfer between the two resonators without back and forth oscillation of the energy.
The effective dissipation rate of the resonator R1 is bounded from above by the coupling strength in the optimal case, $\kappa_\textrm{max,1} = 2 g=\SI{14}{\mega\hertz}$.
In the far detuned case with zero bias voltage, the effective rate of R1 is reduced down to $\kappa_\textrm{min,1} = \kappa_1=\SI{260}{\kilo\hertz}$. 
Thus, the effective damping rate can be tuned by a factor of approximately 50 in our samples.
Larger tuning can be obtained by minimizing the internal losses~\cite{Megrant_short_2012}.
The measurement results are in very good agreement with our  model.
The circuit is based on a QCR, which enables the investigation of the crossover from an underdamped to critically damped and further to overdamped circuit.
In addition to the realization of an EP, the circuit also behaves as a frequency- and voltage-tunable heat sink for quantum electric circuits that can be applied, for example, in quantum information processing for initializing qubits to their ground state by absorbing energy~\cite{Tuorila_2017}.
The tunability of the damping rate enables one to obtain the fastest possible photon absorption allowed by a given coupling coefficient.

In the future, it is interesting to further investigate the EP by modifying the circuit design.
By introducing tunnel junctions to both resonators, one obtains a continuous line of EPs instead of an isolated singularity point.
Incorporating qubits also enables the investigation and utilization of the EP with single energy quanta.
Furthermore, one can investigate a circuit  consisting of resonators with tunable damping realized with a QCR and a tunable coupling realized with a SQUID or a qubit~\cite{Baust_short_2015, Wulschner_short_2016}.
The use of several microwave resonators will result in a more versatile parameter space~\cite{Demange_2012}, and hence yields an interesting platform for studying fundamental physics.
Dynamic encircling of the EP with topological energy transfer~\cite{Dembowski_2001,Doppler_short_2016} can be realized with superconducting resonators in a straightforward manner using standard microwave techniques.
It requires fast tuning of the magnetic field, which can be realized by fabricating a flux bias line on the chip.
Topological energy transfer with microwave pulses may provide an asset for applications in quantum information processing and other quantum technological devices.
In addition, EPs are suitable for investigating $\mathcal{PT}$ symmetry on the level of single microwave photons.
Here, superconducting  circuits provide an attractive architecture owing to the ability to design system parameters yielding, for example, ultra-strong- and deep-strong-coupling regimes~\cite{Quijandria_2018}.
Furthermore, we see EPs as candidates to realize nonreciprocal signal routing beneficial for active quantum circuits~\cite{Metelmann_2018}.

\section*{Acknowledgements}

We thank A. A. Clerk for discussions, and J. Govenius and M. Jenei for assistance.
We acknowledge the provision of facilities and technical support by Aalto University at OtaNano - Micronova Nanofabrication Centre. 
We acknowledge the funding from the European Research Council under  Consolidator Grant No. 681311 (QUESS),
and Marie Sk\l{}odowska-Curie Grant No.~795159,
the Academy of Finland through its Centres of Excellence Program (project Nos.~312300, 312059) and grants (Nos.~265675, 305237, 305306, 308161, 312300, 314302, 316551), 
 the Vilho, Yrj\"o and Kalle V\"ais\"al\"a Foundation,
the Technology Industries of Finland Centennial Foundation,
the Jane and Aatos Erkko Foundation
the Alfred Kordelin Foundation, and
the Emil Aaltonen Foundation.

\section*{Author contributions}

M.P. was  responsible for sample design, fabrication, measurements, data analysis, and  writing the manuscript.
K.Y.T, and J.G. contributed to the sample design. 
L.G. deposited the Nb layer.
J.G., V.S., K.K., and  K.Y.T. contributed to measurements and data analysis.
M.P., K.Y.T, and R.E.L.  developed the fabrication process.
V.V., R.K.,  J.I., and D.H. contributed to the measurements.
M.P.S., A.M., and E.H. contributed to the theory.
M.M. supervised the project.
All authors commented on the manuscript.

\setcounter{table}{0}
\renewcommand{\figurename}{Supplementary Figure}
\renewcommand{\thetable}{S\arabic{table}}
\setcounter{figure}{0}
\renewcommand{\tablename}{Supplementary Table}
\renewcommand{\thefigure}{S\arabic{figure}}

\renewcommand{\theHtable}{Supplement.\thetable}
\renewcommand{\theHfigure}{Supplement.\thefigure}

\newpage

\section*{Methods}

\subsection*{Quantum-circuit refrigerator}

We use a QCR to absorb and emit photons in the resonator R2.
The resonator transition rate from  the occupation number $m$ to $m'$  can be written as~\cite{Silveri_2017}
\begin{equation}
 \Gamma_{m,m'}(V) = M_{mm'}^2 \frac{2 R_\textrm{K}}{R_\textrm{T}} \sum_{\tau=\pm 1} \overset{\rightarrow}{F}[\tau e V + \hbar \omega_2 (m-m')] ,
\end{equation}
where $V=V_\textrm{b}/2$, $R_\textrm{T}$ is the tunneling resistance, $M_{mm'}$ is the corresponding matrix element,  $R_\textrm{K}=h/e^2\approx 25.8$~k$\Omega$  is the von Klitzing constant, and
the normalized rate for forward tunneling is given by
\begin{equation}
\overset{\rightarrow}{F}(E) 
= 
\frac{1}{h} \int_{-\infty}^{\infty}\mathrm{d}E' n_\mathrm{S}(E')[1-f(E',T_\textrm{S})] f(E'-E,T_\textrm{N}),
\end{equation}
where  $f(E,T)=1/\{\exp[E/(k_\textrm{B}T)]+1\}$ is the Fermi--Dirac distribution,  $k_\textrm{B}$ is the Boltzmann constant,
and the density of states in a superconductor can be expressed with the help of the Dynes parameter $\gamma_\textrm{D}$ as 
\begin{equation}
n_\textrm{S}(E) = \left | \textrm{Re}\left( \frac{E/\Delta+i\gamma_\textrm{D}}{\sqrt{(E/\Delta +i\gamma_\textrm{D})^2-1}} \right) \right |.
\label{eq:DOS}
\end{equation}
The matrix element describing the transition can be written in terms of the generalized Laguerre polynomials $L_{n}^{l}(\rho)$ as~\cite{Silveri_2017}
\begin{equation}
 M^2_{m,m'} = 
  \begin{cases}
    e^{-\rho}\rho^{m-m'} \frac{m'!}{m!}[L_{m'}^{m-m'}(\rho)]^2, &  m \geq m', \\ 
    e^{-\rho}\rho^{m'-m} \frac{m!}{m'!}[L_{m'}^{m'-m}(\rho)]^2, &  m < m',
  \end{cases}
\end{equation}
where $\rho = \pi \alpha^2/(\omega_2 C_\textrm{l} x_2 R_\textrm{K})$ is a environmental parameter, where $C_\textrm{l}$ is the capacitance per unit length of the coplanar waveguide, $2 x_2$ is the length of the resonator R2, and the capacitance fraction $\alpha$ is given in terms of the capacitance between the normal-metal island and the center conductor $C_\textrm{N}$, and junction capacitance $C_\textrm{j}$ as $\alpha = C_\textrm{N}/( C_\textrm{N}+4 C_\textrm{j})$.
In the equations above, we have neglected the effects owing to the charging of the normal-metal island since the capacitance of the island is relatively large.
Furthermore, the rates for single-photon transitions can be expressed as~\cite{Silveri_2017}
\begin{equation}
 \begin{aligned}
  \Gamma{_{m,m-1}}
&=&
\kappa_\textrm{QCR} (N+1)m,   \\
  \Gamma{_{m,m+1}} 
&=&
\kappa_\textrm{QCR} N (m+1),
 \end{aligned}
\label{eq:Gamma_single_photon}
\end{equation}
where $\kappa_\textrm{QCR}$ denotes the coupling strength of the QCR,
and the Bose--Einstein distribution at the effective temperature of the electron tunneling, $T_\textrm{QCR}$, is given by  
\begin{equation}
N_\textrm{QCR}=\frac{1}{\exp\left(\frac{\hbar \omega_2 }{k_\textrm{B}T_\textrm{QCR}}\right)-1},
\end{equation}
 where
\begin{equation}
 T_\textrm{QCR} = \frac{\hbar \omega_2}{k_\textrm{B}} 
\left\{ 
\ln\left[
\frac{\sum_{\tau=\pm 1}\overset{\rightarrow}{F}(\tau e V + \hbar \omega_2) }{ \sum_{\tau=\pm 1}\overset{\rightarrow}{F}(\tau e V - \hbar \omega_2)}
\right]
\right\}^{-1}.
\label{eq:T_eff_resonator}
\end{equation}
These equations are derived by defining 
\begin{equation}
 \kappa_\textrm{QCR} = \frac{\Gamma{_{m,m-1}}}{m} -  \frac{\Gamma_{m,m+1}}{m+1}.
\label{eq:kappa_QCR}
\end{equation}

\subsection*{Elastic tunneling in normal-metal--insulator--superconductor junctions}

Typically, the elastic tunneling is the dominating tunneling process.
The electric current through a single NIS junction can be written as~\cite{Giazotto,Pekola_short_2010}
\begin{equation}
I(V)=\frac{1}{e R_\textrm{T}} \int^{\infty}_{0} n_\textrm{S}(E) [f(E-eV, T_\textrm{N})-f(E+eV, T_\textrm{N})] \textrm{d}E,
\label{eq:NIS_current}
\end{equation}
where $T_\textrm{N}$ denotes the normal-metal temperature, and $V$ is the voltage across the junction.
For a symmetric SINIS structure, we apply a voltage $V_\textrm{b}=2V$.
Importantly, this equation has a monotonic dependence on the temperature of the normal metal but only a very weak dependence on the temperature of the superconductor.
Thus, we may use NIS junctions as thermometers measuring the electron temperature of the normal metal.

The tunneling electrons transfer heat through the insulating barrier.
The average power is given by~\cite{Giazotto}
\begin{equation}
P  \!   =   \!   \frac{1}{e^2 R_\textrm{T} }  \!   \int_{-\infty}^{\infty}   \!  \!     n_\textrm{S}(E) (E  -  eV)[f(E  -  eV, T_\textrm{N})  -  f(E, T_\textrm{S})]  \textrm{d}E.
\label{eq:Prefr_NIS_1}
\end{equation}
Based on this equation, we can reduce and increase the temperature of the normal metal.
The applied voltage at the SINIS junction produces a total Joule heating power $P = V_\textrm{b} I$, which is unequally divided between the N and S electrodes.

\subsection*{Quantum mechanical model}

We analyze the temporal evolution of the coupled  resonators following Ref.~\cite{Pierre_2018}.
The Hamiltonian can be written in the rotating wave approximation as
\begin{equation}
 \hat H_\textrm{RWA} 
= \hbar \omega_1 \hat a ^\dagger \hat a 
+ \hbar \omega_2 \hat b ^\dagger \hat b
+ \hbar g( \hat a \hat b^\dagger +  \hat a^\dagger \hat b ).
\end{equation}
The first term describes the energy of the primary resonator R1 with annihilation operator $\hat a$, the second term the energy of R2 with annihilation operator $\hat b$, and the third term describes the coupling between the resonators. 
Here, we have neglected driving.
Furthermore, this equation is valid only for a linear resonator.  
The effects owing to nonlinearity are discussed below.
The dynamics of the system can be obtained from the Lindblad master equation for the density matrix of the coupled system, $\hat{\rho}$, as 
\begin{equation}
 \frac{\textrm{d} \hat{\rho}}{\textrm{d}t}  = 
-\frac{i}{\hbar} [\hat H_\textrm{RWA}, \hat{\rho}] 
+ \kappa_1 \mathcal{L}[\hat a]\hat{\rho} 
+ \kappa_2 \mathcal{L}[\hat b]\hat{\rho}, 
\label{eq:Lindblad}
\end{equation}
where the Lindblad superoperator is given by $\mathcal{L}[\hat x]\hat{\rho}  = \hat x \hat{\rho} \hat x^\dagger -\frac{1}{2} \{\hat x^\dagger \hat x, \hat{\rho} \}$.
We can write the resulting equations of motion as~\cite{Pierre_2018}
\begin{eqnarray}
\frac{\textrm{d} \langle \hat a \rangle}{\textrm{d}t}  &=& -i \omega_1 \langle \hat a \rangle -ig \langle \hat b \rangle -\frac{\kappa_1}{2} \langle \hat a \rangle,  \\
\frac{\textrm{d} \langle \hat b \rangle}{\textrm{d}t}  &=& -i \omega_2 \langle \hat b \rangle -ig \langle \hat a \rangle -\frac{\kappa_2}{2} \langle \hat b \rangle.
\end{eqnarray}
We  define the resonator fields as $\langle \hat a \rangle = A \exp(-i\omega_1 t)$,  $\langle \hat b \rangle = B \exp(-i\omega_1 t)$.
Consequently, the equations assume the form
\begin{eqnarray}
\frac{ \textrm{d}A}{ \textrm{d}t}  &=& -i g B -\frac{\kappa_1}{2}A, \label{eq:time_derivative_A} \\
\frac{ \textrm{d}B}{ \textrm{d}t}  &=& -i \delta B -i g A -\frac{\kappa_2}{2} B,  \label{eq:time_derivative_B}
\end{eqnarray}
where $\delta = \omega_2 - \omega_1$.
These equations can be written in a matrix form as a time-dependent Schr\"odinger equation
\begin{equation}
 \frac{ \textrm{d}}{ \textrm{d}t} \psi = - i H \psi,
\end{equation}
where $\psi = (A, B)^\textrm{T}$, and 
\begin{equation}
 H = 
 \begin{pmatrix}
 -i\frac{\kappa_1}{2}   &  g  \\
  g   & -i\frac{\kappa_2}{2} +\delta
 \end{pmatrix} ,
\end{equation}
as given in Eq.~\eqref{eq:H}.
Here, $H$ is a non-Hermitian Hamiltonian scaled with  $\hbar$.
Equations \eqref{eq:time_derivative_A} and \eqref{eq:time_derivative_B} can also be written as a second-order differential equation,
\begin{equation}
 \frac{ \textrm{d}^2 A}{ \textrm{d}t^2} 
+ \! \left( \! \frac{ \kappa_1 \!  + \!  \kappa_2}{2} +i\delta \right) \! \frac{ \textrm{d} A}{ \textrm{d}t} 
+ \!  \left(g^2 +i\delta \frac{ \kappa_1}{2} +\frac{ \kappa_1  \kappa_2}{4} \right) \! A 
= 0.
 \label{eq:second_order_time_derivative_A}
\end{equation}
When the resonators are tuned into resonance, $\delta = 0$, we can express Eq.~\eqref{eq:second_order_time_derivative_A} as
\begin{equation}
 \frac{\textrm{d}^2 A}{\textrm{d}t^2} 
+ \frac{\kappa_1 + \kappa_2}{2} \frac{\textrm{d}A}{\textrm{d}t} 
+ \left(g^2 + \frac{\kappa_1  \kappa_2}{4} \right) A
= 0.
\end{equation}
This equation describes a damped harmonic oscillator, where the energy is transferred between the resonators R1 and R2 at an angular frequency $\sqrt{g^2 + \kappa_1  \kappa_2/4 }$.
Due to the asymmetric damping rates in the two resonators, the total dissipation rate of the system is time-dependent and reaches its maximum value when the excitations are in R2.
The damping ratio is given by
\begin{equation}
 \xi  = \frac{\kappa_1 + \kappa_2}{2\sqrt{4g^2 +\kappa_1 \kappa_2}}.
\end{equation}
Here, $\kappa_2$ is a function of voltage $V_\textrm{b}$, which allows us to examine the transition from an underdamped system, $\xi < 1$, through critical damping, $\xi = 1$, to an overdamped system, $\xi > 1$.
Critical damping is obtained when $|\kappa_2-\kappa_1| = 4g$.
The total damping rate of R1 is given by  $\kappa_1=  \kappa_\textrm{int,1} +\kappa_\textrm{ext}$, and of R2 by $\kappa_2(V_\textrm{b}) = \kappa_\textrm{int,2} + \kappa_\textrm{QCR}(V_\textrm{b}) + \kappa_\textrm{r,2}(V_\textrm{b})$, where $\kappa_\textrm{int,1/2}$ denote the internal losses, $\kappa_\textrm{ext}$ the losses to the external measurement circuit, $\kappa_\textrm{QCR}(V_\textrm{b})$ the photon-assisted tunneling in Eq.~\eqref{eq:kappa_QCR}, and $\kappa_\textrm{r,2}(V_\textrm{b})$ the residual voltage-dependent losses in R2.
In our samples  $\kappa_\textrm{2}(V_\textrm{b}) \gg  \kappa_\textrm{1}$,  and  $g \gg  \kappa_\textrm{int,1} \gg \kappa_\textrm{ext}$, as discussed below.
Therefore, we obtain an approximate condition for the critical damping as 
\begin{equation}
 \kappa_2 = 4 g.
\label{eq:kappa_2_critical_condition}
\end{equation}
The critical damping, which corresponds to the EP, is obtained at $eV_\textrm{b}/(2\Delta) \approx 1$ where the photon number remains low, and therefore, the slight nonlinearity caused by the SQUID is of negligible importance.
However, at  $eV_\textrm{b}/(2\Delta) > 1$, the QCR generates thermal photons that result in photon-number-dependent losses, as discussed below.

\subsection*{Sample parameters}

The main  parameters for the samples are summarized in Supplementary Table~\ref{tab:simulation_parameters}.
The coupling strength between the resonators can be estimated as~\cite{Jones_4} $g=C_\textrm{C} V_1 V_2 / \hbar \approx 2\pi \times \SI{7.2}{\mega\hertz}$, where the voltages are given by $V_i=\sqrt{\hbar \omega_1 /(2 x_i C_\textrm{l})}$, $i=1,2$,  the angular frequency of the second mode of the resonator R1 is $\omega_1/(2\pi) \approx \SI{5.223}{\giga\hertz}$ for Samples A and B, and $C_\textrm{l}$ is the capacitance per unit length.
Consequently, the critical damping is obtained with $\kappa_2 = 4g  \approx 2\pi \times \SI{29}{\mega\hertz}$.
The external quality factor corresponding to the leakage from the resonator R1 to the transmission line through the capacitances $C_\textrm{TL}$ is given by~\cite{Goppl_short_2008} $Q_\textrm{ext}= 2 x_1 C_\textrm{l} /(4 Z_\textrm{L} \omega_1 C_\textrm{TL}^2) \approx 9\times 10^5$.
Consequently, the corresponding damping rate is $\kappa_\textrm{ext} = \omega_1 / Q_\textrm{ext} \approx 2\pi \times \SI{6}{\kilo\hertz}$.
The loaded quality factor of the second mode of R1 is approximately $Q_\textrm{L}=2\times10^4$, when the resonators are far detuned. 
Thus, the internal losses in R1 dominate over the losses to the transmission line, $Q_\textrm{int}\approx Q_\textrm{L}$.
Furthermore, we obtain the damping rate $\kappa_1 = \omega_1 / Q_\textrm{L} \approx 2\pi \times \SI{300}{\kilo\hertz}$.
The real part of the complex wave propagation coefficient, $\gamma = \alpha + i \beta$, describes the damping in the waveguide, and it can be calculated as~\cite{Goppl_short_2008} $\alpha = n_\textrm{m} \pi /(4 x_1 Q_\textrm{int})\approx \SI{7e-3}{\per\metre}$, where $n_\textrm{m}$ is the mode number with $n_\textrm{m}=2$ denoting the first excited mode.
The internal losses without the photon-assisted tunneling in the QCR are somewhat higher in the resonator R2 than in R1 since the design and fabrication of the QCR and the SQUID have not been optimized for  low loss rates.
The internal loss rate for R2 can be extracted at zero detuning and zero bias voltage  from the saturation level of extracted $\kappa_2$ values since $\kappa_\textrm{int,1} \gg \kappa_2(0)$, and hence the losses in R2 dominate over those in R1.
We obtain from the circuit model the internal loss rate for R2 as $\kappa_\textrm{int,2}\approx\kappa_2(0) = 2\pi \times \SI{16}{\mega\hertz}$.

The photon number inside R1 when R2 is far detuned can be estimated as~\cite{Oelsner_short_2017} $n= 4 \Omega_\textrm{d}^2/\kappa_1^2 \approx 10$, where the driving strength is given by $\Omega_\textrm{d}=C_\textrm{TL} V_\textrm{in} V_1 / \hbar $, the input voltage is obtained from the input power as $V_\textrm{in}=\sqrt{P_\textrm{in} Z_\textrm{L}}$, and the input power is $P_\textrm{in} \approx -115$~dBm.
The input power is $-100$~dBm for Sample~A in Fig.~\ref{fig:meas_sim_crossing_selected} and Supplementary Fig.~\ref{fig:meas_sim_crossing_sampleA}, and $-115$~dBm for Sample~B in Supplementary Fig.~\ref{fig:meas_sim_crossing_sampleB}.
We also measure the resonators at different power levels.
When R1 and R2 are in resonance at $V_\textrm{b}=0$, the total photon number is approximately equally divided between the resonators if $\kappa_1\approx\kappa_2$.
However, in our samples $\kappa_1 < \kappa_2$, especially at $V_\textrm{b} > 2\Delta/e$, and therefore the number of coherent photons is lower in R2 than in R1.
When the $Q$ factor of the resonator is reduced to 200, which is of the order of the critical damping,  photon numbers close to unity are obtained with an input power  $P_\textrm{in} \approx -85$~dBm.

\subsection*{Residual losses in the resonator R2}

We  attribute the residual voltage-dependent losses to dephasing, and to  dissipation sources such as quasiparticle generation in the superconductors and resistive losses in the normal metal.
Firstly, the resonator R2 is slightly nonlinear owing to the SQUID, and hence, an increasing incoherent photon number results in dephasing. 
Dephasing can be added in Eq.~\eqref{eq:Lindblad}  with a term $\kappa_\phi \mathcal{L}[\hat b^\dagger\hat b]\hat{\rho}$, where the dephasing rate $\kappa_\phi$ depends on the number of thermal photons in the resonator.
Similarly, in the case of superconducting qubits, the dephasing can be written as  $\kappa_\phi \mathcal{L}[\hat{\sigma}_z]\hat{\rho}$, where $\hat{\sigma}_z$ is a Pauli operator.
The factor $\kappa_\phi$ causes a similar effect as $\kappa_2$ in Eqs.~\eqref{eq:Lindblad}--\eqref{eq:kappa_2_critical_condition} although it does not decrease the total photon number in the resonators.
The photon number variance  for a thermal state is of the form~\cite{Clerk_2007,Goetz_short_2017} $n(n+1)$, and therefore, thermal photons cause more dephasing than the coherent photons with a variance of $n$, where $n$ is the average photon number.
Consequently, we assume that $\kappa_\phi = \omega_\phi n (n+1)$, where $\omega_\phi$ is a proportionality coefficient.
Furthermore, as discussed above, the number of the coherent photons is low in R2 due to the relatively high loss rate.
The steady-state photon number in the resonator can be estimated as~\cite{Silveri_2017}
\begin{equation}
 n = \frac{\kappa_\textrm{QCR} N_\textrm{QCR} }{\kappa_\textrm{QCR} + \kappa_\textrm{int,2}},
 \label{eq:photon_number_in_R2}
\end{equation}
where we assume that the photon number of the  effective bath, to which R2 is coupled through $ \kappa_\textrm{int,2}$, vanishes owing to the very low cryostat temperatures of approximately $\SI{10}{\milli\kelvin}$.
The photon number depends linearly on the bias voltage at voltages above the superconductor energy gap, as shown in Supplementary Fig.~\ref{fig:transition_rate_parameters}.

Secondly, we take the quasiparticle losses into account.
The critical temperature of Nb is approximately $\SI{9}{\kelvin}$, and therefore, the quasiparticle density remains low in it.
However, the critical temperature of Al approximately $\SI{1.2}{\kelvin}$, which enables higher quasiparticle density than in Nb.
We observe a decrease in the critical current of the SQUID, which indicates increased temperature in the Al leads of the SQUID, and hence heat dissipation.
The quasiparticle loss rate~\cite{Barends_short_2011,Oneil_2012} $\kappa_\textrm{qp} \propto n_\textrm{qp} \propto \sqrt{P}$, where $P$ is the absorbed power.
The Al leads at the NIS junctions receive  half of the Joule power $P= IV_\textrm{b}$ at high voltages, whereas the other half is absorbed to the normal metal.
Thus, the power is quadratic in voltage, which is linear in the estimated photon number.
Therefore,  the expected quasiparticle losses are linear in photon number, $\kappa_\textrm{qp} = \omega_\textrm{qp} n$, where $\omega_\textrm{qp}$ is a proportionality coefficient.
The dc power dissipated in the junctions is substantially higher than the microwave input power.
At $eV_\textrm{b}/(2\Delta) = 2$, the dc power is approximately $\SI{30}{\pico\watt}$ compared to a microwave power of $-100$~$\textrm{dBm} =\SI{0.1}{\pico\watt}$.
The normal metal in the QCR acts as an effective quasiparticle trap~\cite{Oneil_2012} minimizing the quasiparticle losses.
Some fraction of the power dissipated at the QCR  leaks to the SQUID.

There is an approximately 10-$\mu$m-long section of normal metal between the actual Nb resonator and the NIS junctions, which may cause some losses.
The loss rate at the resistor depends on the current profile of the microwave mode, which  can depend on the voltage $V_\textrm{b}$.
Nevertheless, we assume these losses to be small due to the QCR being at the end of the resonator.
Furthermore, there is a layer of superconducting Al below the normal metal due to the shadow evaporation technique, which decreases the current in the resistor, and hence also the resistive losses.
The very weak resistive losses are quadratic in the voltage amplitude of the microwave resonator which is linear in photon number.
Thus, it can be approximated as  $\kappa_\textrm{res} = \omega_\textrm{res} n$ with a proportionality coefficient $\omega_\textrm{res}$.

Consequently, the total voltage-dependent losses in R2 including the  dephasing, quasiparticle losses in the superconductors and the resistive losses  are given by
\begin{equation}
 \kappa_\textrm{r,2} = \kappa_\phi + \kappa_\textrm{qp} +\kappa_\textrm{res} = \kappa_\phi n(n+1) + \kappa_\textrm{qp} n + \kappa_\textrm{res} n.
\end{equation}
The quasiparticle and resistive losses are expected to be very weak, as described above, but a small contribution cannot be excluded.
Nevertheless, we expect the dephasing  to dominate over the quasiparticle and resistive losses.
Therefore, in the numerical analysis, we take the photon-number-dependent losses into account as
\begin{equation}
  \kappa_\textrm{r,2} =   \omega_\textrm{r,tot} n (n+1),
\label{eq:kappa_residual_simple}
\end{equation}
with only one fitting parameter $ \omega_\textrm{r,tot}$ effectively describing the different loss methods discussed above.
From the experimental damping rates of the dissipative resonator R2, we extract the the coefficient $ \omega_\textrm{r,tot} \approx 2\pi\times \SI{22}{\mega\hertz}$.
The good agreement with the experimental damping rate $\kappa_2$ and the model with the quadratic residual losses $\kappa_\textrm{r,2}$ in Fig.~\ref{fig:transition_rates}(a) gives further support for the approximation in Eq.~\eqref{eq:kappa_residual_simple}.
We do not take this loss rate into account in Eq.~\eqref{eq:photon_number_in_R2} for simplicity, and also due to the fact that pure dephasing does not decrease the photon number.

The odd modes of R1 do not show flux dependence as expected due to the voltage node at the coupling capacitor.
However, they do show some dependence on the voltage $V_\textrm{b}$.
Similar dependence can be observed also for the even modes at $\Phi/\Phi_0 = 0.5$ where the inductance of the SQUID ideally vanishes and  thus decouples the QCR from the resonator R1.
We attribute this observation to unintentional asymmetry of the sample. 
Furthermore, the QCR may be weakly coupled to the input and output microwave fields through some spurious mode of the sample holder.
The very broad resonance at high bias voltages enables the coupling to the spurious modes.
We note that the spurious modes may be partially responsible for the $\kappa_\textrm{r,2}$.
However, we do not quantitatively model these losses. 
Instead, they are effectively included in the parameter $\omega_\textrm{r,tot}$ in Eq.~\eqref{eq:kappa_residual_simple}.

\subsection*{Full model for $\kappa_2$ and $\kappa_\textrm{eff}$}

The parameters  $\kappa_2$ and $\kappa_\textrm{eff}$ are obtained as follows.
First, we extract the effective resistance corresponding to the QCR by fitting the classical circuit model to the experimentally obtained scattering parameter $S_{21}$ using a least-squares algorithm.
Second, we calculate the quality factor of the resonator R2, $Q_\textrm{R2}$, for the obtained effective resistance, as discussed below.
The coupling rate is related to the quality factor as $\kappa_2 = \omega_2/Q_\textrm{R2}$. 
The full model denoted by the line in Fig.~\ref{fig:transition_rates}(a) is obtained by fitting 
\begin{equation}
 \kappa_2(V_\textrm{b}) = \kappa_\textrm{QCR}(V_\textrm{b}) +  \kappa_\textrm{r,2}(V_\textrm{b}) + \kappa_\textrm{int,2}
\end{equation}
to the experimental transition rates according to Eqs.~\eqref{eq:kappa_QCR}, \eqref{eq:photon_number_in_R2}, and \eqref{eq:kappa_residual_simple}.
Here, we use $\omega_\textrm{r,tot}$ as the only fitting parameter since we fix $\kappa_\textrm{int,2}$ to the saturation value at zero bias, as discussed above.

Subsequently, we may proceed to the effective damping rates $\kappa_\textrm{eff} = -2 \textrm{Im}(\lambda_\pm)$, which can be obtained from $\kappa_2$ with the help of  Eq.~\eqref{eq:lambda}. 
The damping rates above the critical damping, $\kappa_2 > 4g$, result in the two branches at bias voltages $V_\textrm{b} \gtrsim 2\Delta/e$.
The lines in Fig.~\ref{fig:transition_rates}(b) are obtained similarly from  Eq.~\eqref{eq:lambda}.

\subsection*{Classical circuit model}

To simulate the scattering parameter $S_{21}$, we use a classical circuit model similar to the one presented in Ref.~\cite{Partanen_short_2018}.
We  analyze the samples using standard microwave circuit analysis~\cite{Pozar}.
The input impedance of the resonator R2 is
\begin{equation}
 Z_\textrm{R2} \! = \! 
 Z_\textrm{C} + 
 \frac{Z_0 \!  \left\{ \!   Z_\textrm{S} + Z_0 \tanh(\gamma x_2)   \! +  \! \frac{Z_0 \left[ R_\textrm{eff} +  Z_0 \tanh(\gamma x_2) \right]}{Z_0  + R_\textrm{eff} \tanh(\gamma x_2)} \right\}
}{
Z_0 +  \tanh(\gamma x_2)  \left\{ Z_\textrm{S}  + \frac{Z_0 [ R_\textrm{eff} +  Z_0 \tanh(\gamma x_2) ]}{Z_0  +  R_\textrm{eff} \tanh(\gamma x_2)}  \right\}
},
\end{equation}
where the impedance of the SQUID and the capacitors between the SQUID and the center conductor is given by $Z_\textrm{S}=i\omega L_\textrm{S} + 2/(i\omega C_\textrm{S})$, the impedance of the coupling capacitor between the resonators by $Z_\textrm{C}= 1/(i \omega C_\textrm{C})$, $\gamma$ is the complex propagation coefficient discussed above,
and the terminating impedance consisting of the effective resistance of the NIS junctions and the capacitor between the normal-metal island and the center conductor is modeled as an effective resistor with resistance  $R_\textrm{eff}$. 
The inductance of the  SQUID is calculated as
$ L_\textrm{S}(\Phi) = \Phi_0 /[2 \pi I_0 |\cos(\pi \Phi / \Phi_0)|],$
where the maximum supercurrent through the SQUID is $I_0$, and the flux quantum is   $\Phi_0 = h/(2e)$.

The scattering parameter $S_{21}$ describing the voltage transmission from Port~1 to Port~2 can be calculated using the transmission matrix method as~\cite{Pozar}
\begin{equation}
S_{21}=\frac{2}{ A_\textrm{m}  + B_\textrm{m}/Z_\textrm{L}  + C_\textrm{m} Z_\textrm{L} +  D_\textrm{m}},
\end{equation}
where $Z_\textrm{L}$ is the characteristic impedance of the external measurement cables, and 
\begin{equation}
 \begin{pmatrix}
 A_\textrm{m}   &B_\textrm{m}  \\
 C_\textrm{m}   &D_\textrm{m}
 \end{pmatrix} 
= M_1 M_2 M_3 M_2 M_1,
\label{eq:ABCD}
\end{equation}
with
\begin{eqnarray}
 M_1 &=& 
\begin{pmatrix}
 1 &\frac{1}{i \omega C_\textrm{TL}}\\
0 &1
 \end{pmatrix} ,
\\
 M_2 &=&
 \begin{pmatrix}
 \cosh(\gamma x_1) & Z_0 \sinh(\gamma x_1) \\
\frac{1}{Z_0} \sinh(\gamma x_1)  & \cosh(\gamma x_1) 
 \end{pmatrix} ,
\\
 M_3 &=&
 \begin{pmatrix}
 1 &0\\
\frac{1}{Z_\textrm{R2}} &1
 \end{pmatrix} .
\end{eqnarray}
We analyze the losses in the resonator R2 also in the absence of coupling to R1. 
In particular, we omit matrices $M_1$ and $M_2$ from Eq.~\eqref{eq:ABCD}.
The resonator R2 causes a dip in the amplitude of the transmission coefficient $S_{21}$, whereas R1 causes a peak.
The quality factor can be estimated directly from the ratio of the center frequency and the width of the peak or dip.
Alternatively, more advanced methods can be used~\cite{Petersan_1998}.

\subsection*{Sample fabrication}

The samples are fabricated on a  Si wafer with a thickness of $\SI{500}{\micro\meter}$ and a diameter of $\SI{100}{mm}$. 
First, a 300-nm-thick layer of SiO$_2$ is thermally grown on the wafer with resistivity $\rho>\SI{10}{\kilo\ohm\centi\meter}$.
Subsequently, a 200-nm-thick layer of Nb is sputtered on top of the oxide.
The resonators are patterned on the Nb layer with optical lithography and reactive ion etching.
We cover the complete wafer with a 40-nm-thick layer of Al$_2$O$_3$ fabricated using atomic layer deposition.
This oxide layer serves as an insulating barrier in the parallel plate capacitors and separates the QCR lines from the ground plane.
The nanostructures are defined using electron beam lithography and two-angle shadow evaporation followed by  a lift-off process.
The SQUID consists of two Al layers with thicknesses of $\SI{40}{\nano\metre}$ each.
The first Al layer is oxidized \emph{in situ} in the evaporation chamber at $\SI{1.0}{\milli\bar}$ for $\SI{5}{\min}$.
The SINIS junctions consist of Al ($\SI{40}{\nano\metre}$) and Cu ($\SI{40}{\nano\metre}$), and the Al layer is similarly oxidized as in the SQUID.
The shadow evaporation technique  results in overlapping metal layers.

\subsection*{Measurement setup}

The measurement setup is schematically presented in Fig.~\ref{fig:measurement_setup}.
The samples are measured in a commercial dry-dilution refrigerator with a base temperature of approximately 10~mK.
The scattering parameters are measured with a vector network analyzer which contains both the microwave source and the detector. 
The microwave signal is attenuated at different temperature stages to avoid heat leakage from  higher temperatures to the sample.
We employ amplifiers at $\SI{4}{K}$ and at room temperature.
The NIS junctions are controlled by applying a bias voltage or current through continuous thermocoax cables from room temperature down to the base temperature.
Magnetic flux for the SQUID is produced using a superconducting coil with a bias current.

\subsection*{Normalization of scattering parameters}

All measured scattering parameters $S_{21}$ are normalized.
Initially, we normalize the phase winding originating from the electrical delay $\tau \approx \SI{50}{\nano\second}$ in  the measurement setup outside the sample by multiplying with $\exp(i\omega \tau)$.
Consequently, the resonance produces a circle on the complex plane  as the frequency is increased over the resonance.
We transform this circle to its canonical position where $\max|S_{21}|$  is on the positive real axis and the circle intersects the origin.
Finally, we normalize the amplitude to unity by dividing with $\max|S_{21}|$.

\begin{table*}[h]
 \begin{center}
 \caption{ Parameters. 
The parameters for Sample~B that differ from those for Sample~A are given in parenthesis. 
See Methods and Supplementary Fig.~\ref{fig:measurement_setup}(b) for details.
The resonance frequency of the first excited mode of the resonator R1, $f_1$, is a measured value, 
the characteristic impedances of the transmission lines in the resonator, $Z_0$, and in the external measurement circuit, $Z_\textrm{L}$, are nominal values.
The lengths of the resonator sections $x_1$ and $x_2$ are design values, and the effective resistance is calculated as~\cite{Goppl_short_2008} $\sqrt{\varepsilon_\textrm{eff}} = c/(2 f_1 x_1 )$, where $c$ is the speed of light in vacuum.
We obtain the values for the capacitance per unit length $C_\textrm{l}$, and the capacitance $C_\textrm{TL}$ from finite element method (FEM) simulations.
The capacitance $C_\textrm{C}$ is obtained by fitting the circuit model to the measured scattering parameter $|S_{21}|$ in good agreement with FEM simulations, and the capacitances $C_\textrm{N}$, $C_\textrm{S}$ and $C_\textrm{j}$  are calculated using a parallel-plate model.
The coupling strength $g$ is obtained from  $C_\textrm{C}$.
The loaded quality factor of the first exited mode of R1 $Q_\textrm{int,1}$ and the tunneling resistance $R_\textrm{T}$ are  measured values, and the Dynes parameter $\gamma_\textrm{D}$ is estimated as the ratio of the asymptotic resistance and the resistance in the superconductor gap.
The critical current at zero-bias $I_\textrm{c,0}$ is given by the flux corresponding to the crossing of the modes in the circuit model in good agreement with a control sample with slightly smaller junction area and a critical current of approximately 200~nA.
The damping rate $\kappa_1$ is given by the ratio $\omega_1/Q_\textrm{int,1}$, and the damping rate $\kappa_\textrm{int,2}$ is extracted from the saturation value of $\kappa_2$ at zero bias.
The proportionality coefficient for the residual losses $\omega_\textrm{r,tot}$ is a fitted value.
}
 \begin{tabular}{ c | c  }
\hline
Parameter &  Value \\
\hline
\hline
$f_1$ &  5.223 GHz \\ 
$Z_0$ & 50 $\Omega$ \\
$Z_\textrm{L}$ & 50 $\Omega$ \\
$x_1$ & 12 mm \\
$x_2$ & 6.0  (6.5)  mm\\
$\varepsilon_\textrm{eff}$ & 5.73 \\
$C_\textrm{l}$ & 155 pF/m \\
$C_\textrm{TL}$ & 0.8 fF \\
$C_\textrm{C}$ & 3.8 fF \\
$C_\textrm{N}$ & 98 fF \\
$C_\textrm{S}$ &  460 fF \\
$C_\textrm{j}$ & 6.2 fF \\
$g/(2\pi)$ & 7.2~MHz \\
$Q_\textrm{int,1}$ & $2.7\times 10^4$  $(2.0\times 10^4)$ \\
$R_\textrm{T}$ & 8.4  (9.5) k$\Omega$ \\
$\gamma_\textrm{D}$ & $1\times10^{-4}$ \\
$I_\textrm{c,0}$ & 340  (300)~nA\\
$\kappa_\textrm{1}/(2\pi)$ & 190 (260)~kHz \\ 
$\kappa_\textrm{int,2}/(2\pi)$ & 16~MHz \\
$\omega_\textrm{r,tot}/(2\pi)$ & 22~MHz \\
\hline
\end{tabular}
 \label{tab:simulation_parameters}
 \end{center}
\end{table*}

\begin{figure*}[h]
\centering
\includegraphics[width=160mm]{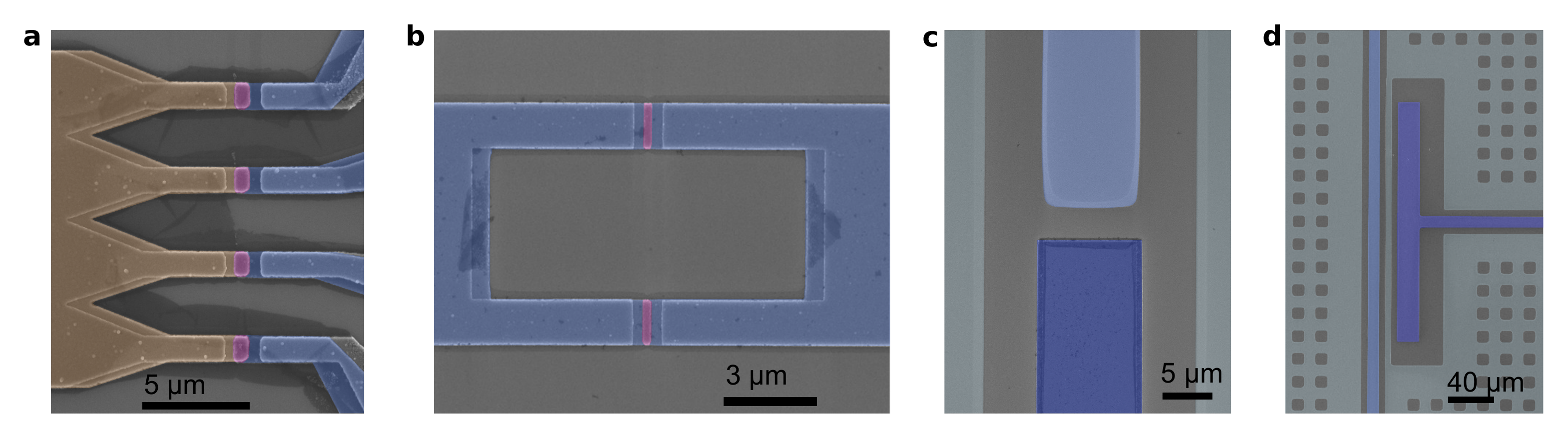}
\caption{False-color scanning electron micrographs of the sample.
(a)~Normal-metal island with four NIS junctions highlighted in purple.
(b)~SQUID with two Josephson junctions highlighted in purple.
(c)~Coupling capacitance between R1 (light blue) and an external port (dark blue).
(d)~Coupling capacitance between R1 (light blue) and R2 (dark blue).
}
\label{fig:sample_details}
\end{figure*}

\begin{figure*}[h]
\centering
\includegraphics[width=105mm]{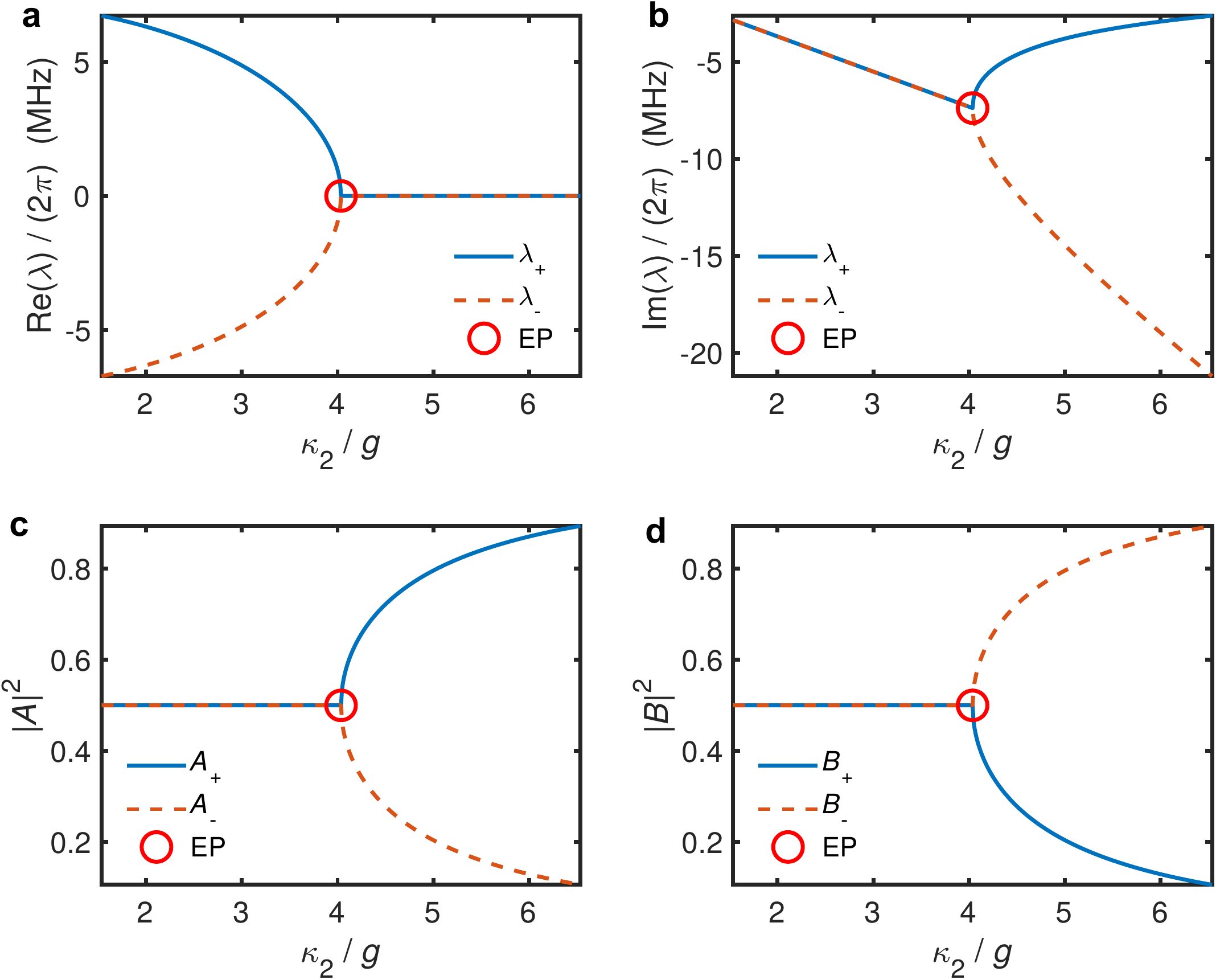}
\caption{Eigenvalues and eigenvectors of the effective Hamiltonian.
(a)~The real part of the eigenvalues corresponding to the frequency shifts from the bare resonator R1 mode frequency as a function of the decay rate $\kappa_2$ at zero detuning. 
(b)~The imaginary part of the eigenvalues corresponding to negative decay rates.
(c)~The squared absolute value of the eigenvector component corresponding to the resonator R1. 
Here, the amplitude of the eigenvector $\Psi_\pm = (A_\pm, B_\pm)^\textrm{T}$ is normalized to unity.
(d) As (c) but for the eigenvector component corresponding to R2.
}
\label{fig:eigenvalues_eigenvectors}
\end{figure*}

\begin{figure*}[h]
\centering
\includegraphics[width=140mm]{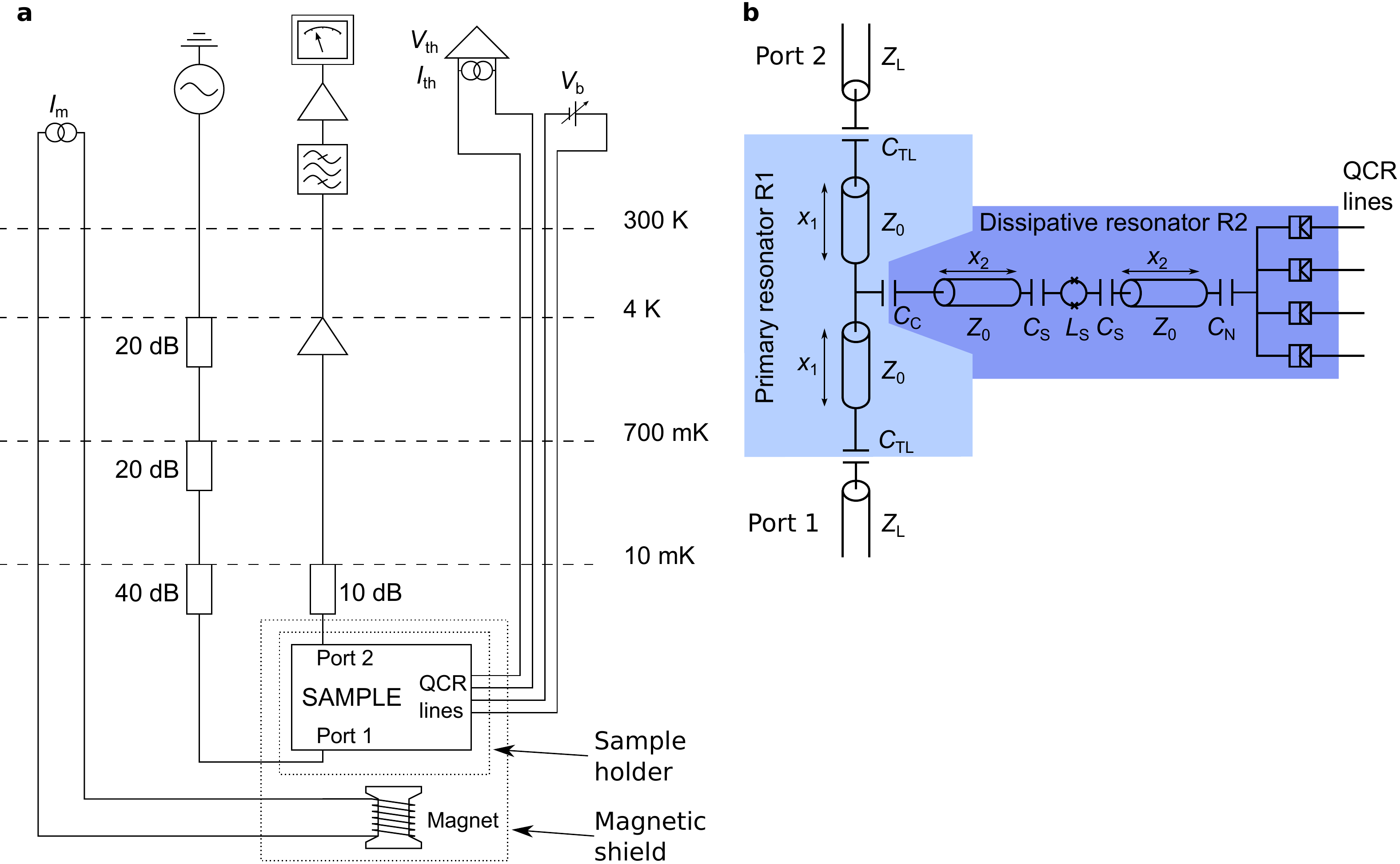}
\caption{Measurement setup and circuit diagram of the sample.
(a)~Simplified measurement setup showing the attenuators, and amplifiers at different temperatures.
We measure the sample response to microwave signal from Port~1 to Port~2.
Magnetic field for the SQUID is generated using a coil with current $I_\textrm{m}$.
A bias voltage $V_\textrm{b}$, and bias current for thermometry $I_\textrm{th}$ are applied to the NIS junctions.
The temperature of the normal metal is deduced from voltage $V_\textrm{th}$ measured with an applied bias current $I_\textrm{th}$.
(b)~Sample structure presented as an electrical circuit diagram.
The transmission lines of the resonators have characteristic impedances $Z_0$, and the external transmission lines $Z_\textrm{L}$.
The sections of the resonators have lengths $x_1$ and $x_2$. 
The capacitances at the external ports are denoted by $C_\textrm{TL}$, between the resonators by $C_\textrm{C}$, between the SQUID with inductance $L_\textrm{S}$ and the center conductor of the transmission line by $C_\textrm{S}$, and between the normal-metal island and the center conductor by $C_\textrm{N}$. 
}
\label{fig:measurement_setup}
\end{figure*}

\begin{figure}[h]
\centering
\includegraphics[width=88mm]{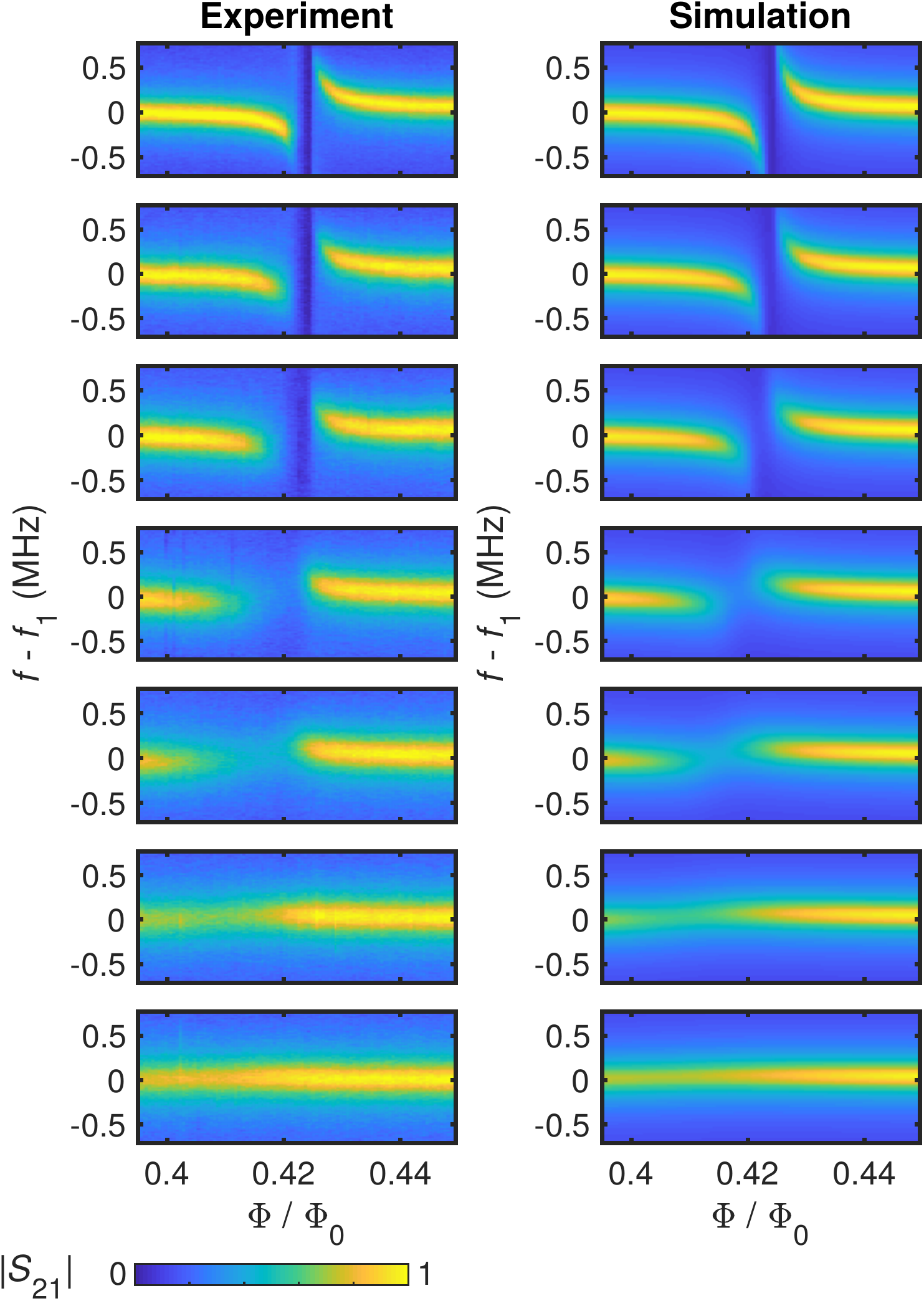}
\caption{ Measured and simulated scattering parameter $|S_{21}|$ for  Sample~A as a function of frequency and flux for different bias voltages.
The bias voltages from top to bottom are 
$eV_\textrm{b}/(2\Delta) =  $   
    0.0,
    1.0,
    1.1,
    1.4,
    1.7,
    2.5, and
    3.5.
Maximum value in each panel is normalized separately to unity. 
The input power at Port~1 is approximately $-100$~dBm.
}
\label{fig:meas_sim_crossing_sampleA}
\end{figure}

\begin{figure}[h]
\centering
\includegraphics[width=88mm]{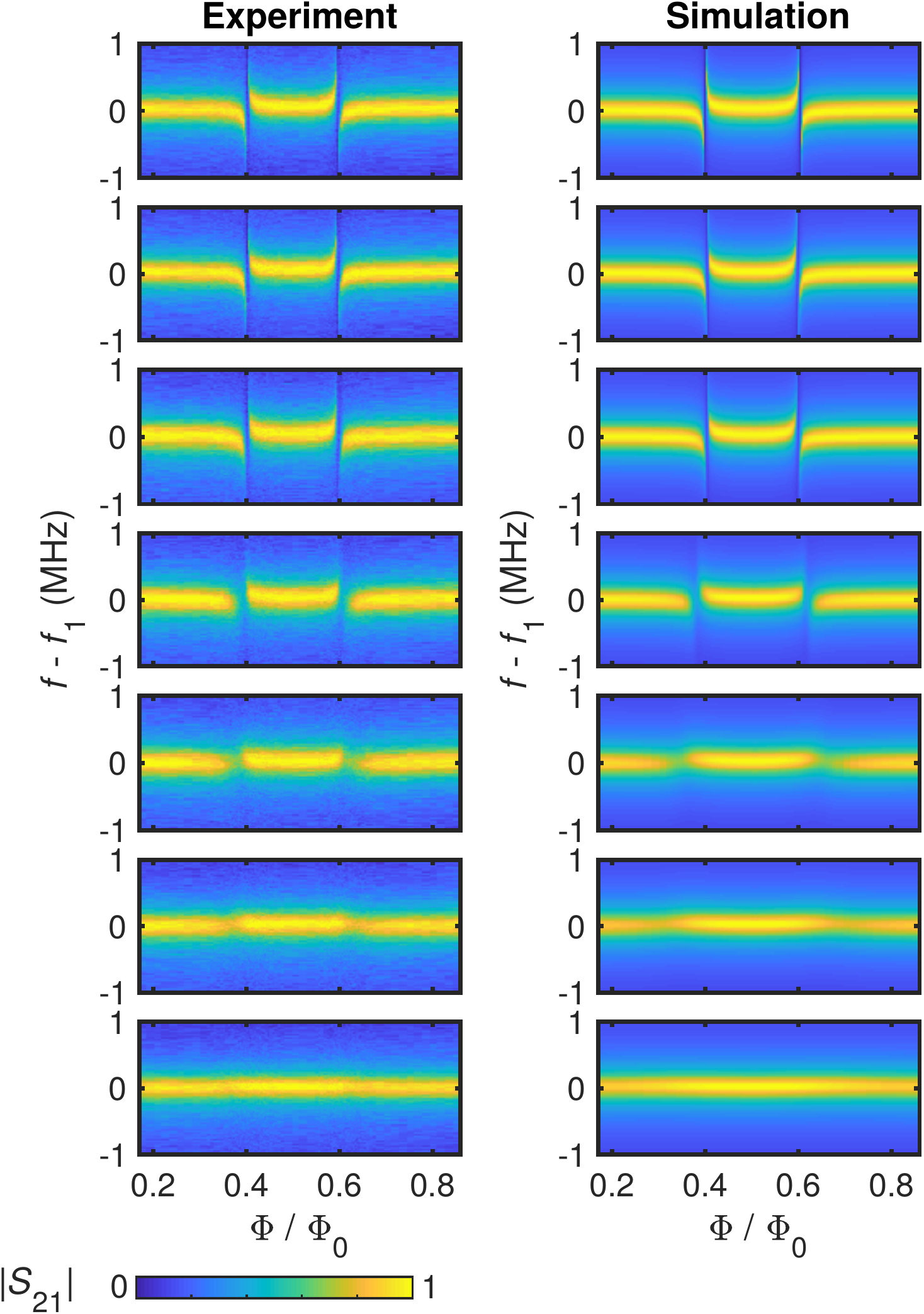}
\caption{ Measured and simulated scattering parameter $|S_{21}|$ for  Sample~B as a function of frequency and flux for different bias voltages.
The bias voltages from top to bottom are 
$eV_\textrm{b}/(2\Delta) =  $   
    0.0,
    0.6,
    1.1,
    1.6,
    2.4,
    3.7, and
    6.2.
The maximum value in each panel is normalized to unity.
The input power at Port~1 is approximately $-115$~dBm.
}
\label{fig:meas_sim_crossing_sampleB}
\end{figure}

\begin{figure*}[h]
\centering
\includegraphics[width=130mm]{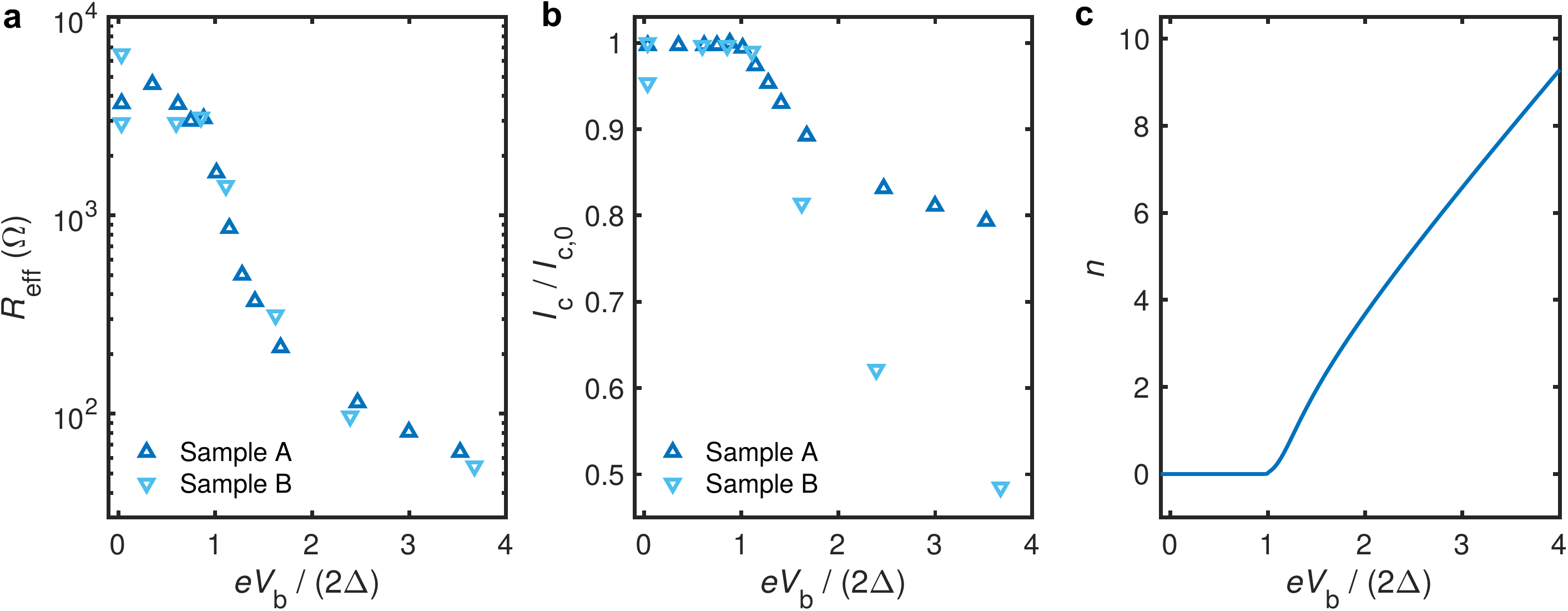}
\caption{ Extracted parameters for the transition rates for Samples~A and~B. 
(a)~The effective resistance in the circuit model as a function of the bias voltage.
(b)~Critical current $I_\textrm{c}$ normalized with the maximum critical current $I_\textrm{c,0}$.
(c)~Estimated average photon number in the dissipative resonator  R2.
}
\label{fig:transition_rate_parameters}
\end{figure*}

\begin{figure*}[h]
\centering
\includegraphics[width=130mm]{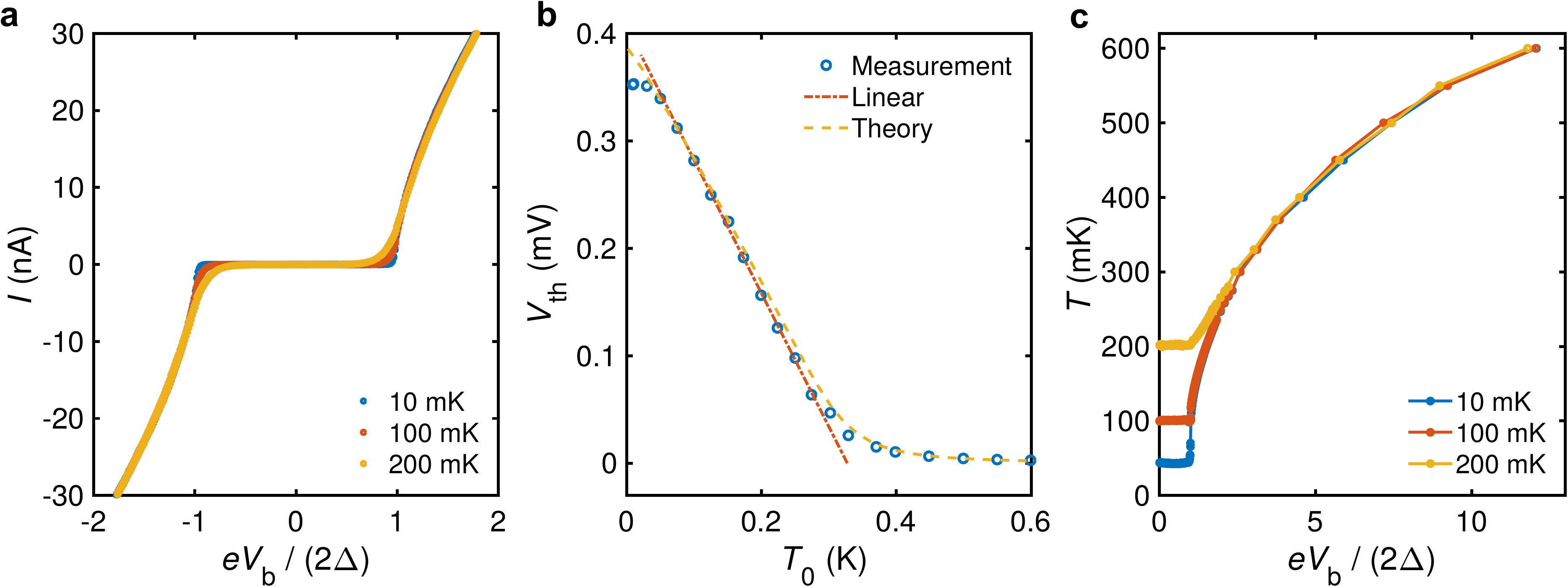}
\caption{ Current--voltage characteristics and temperature for Sample~B.
(a)~Electric current through a SINIS junction as a function of bias voltage at different bath temperatures.
(b)~Measured thermometer voltage $V_\textrm{th}$ as a function of bath temperature  $T_0$ at fixed bias current  $I_\textrm{th}=\SI{17}{\pico\ampere}$, and  $V_\textrm{b}=0$.   
The theoretical curve is calculated using Eq.~\eqref{eq:NIS_current}. 
We extract the electron temperatures of the normal-metal island using a linear voltage-to-temperature conversion below 300~mK, and above that we extract the temperatures from the voltages corresponding to the different experimental bath temperature points. At high temperatures, the low sensitivity  reduces the reliability of the extracted island temperatures.
(c)~Electron temperature of the normal-metal island as a function of bias voltage at different bath temperatures.
}
\label{fig:IV_temperature_curves}
\end{figure*}

\begin{figure*}[h]
\centering
\includegraphics[width=90mm]{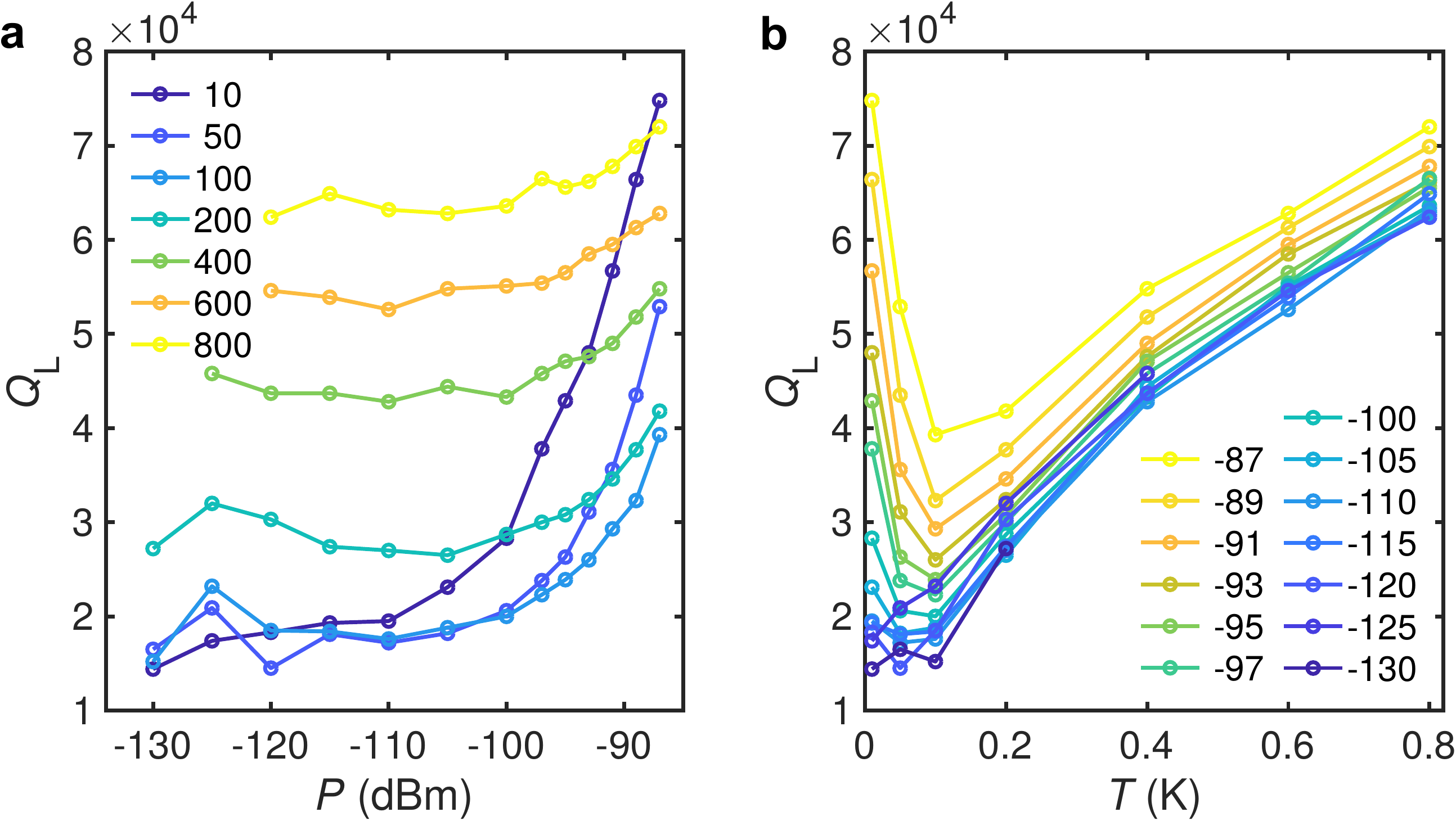}
\caption{ Measured quality factor of Sample~B.
(a)~$Q_\textrm{L}$  as a function of power at different bath temperatures as indicated in mK. The flux is $\Phi_0/2$.
(b)~As (b) but the data is presented as a function of the bath temperature at different powers as indicated in dBm.
The lines are guides for the eye.
}
\label{fig:meas_Qfactor}
\end{figure*}

\clearpage

\end{document}